\documentclass[article]{JHEP}
\usepackage{amsmath,epsfig}
\usepackage{amssymb,amsfonts}
\usepackage{latexsym}

\usepackage{epsfig}

\relax
\renewcommand{\theequation}{\arabic{section}.\arabic{equation}}

\def\be{\begin{equation}}
\def\ee{\end{equation}}

\newcommand{\rc}{\nonumber\\}

\newcommand{\bear}{\begin{eqnarray}}

\newcommand{\eear}{\end{eqnarray}}
\def\hre#1#2{\href{http://arxiv.org/abs/#1/#2}{[ArXiv:#1/#2]}}
\def\hspi#1#2{\href{http://www.slac.stanford.edu/spires/find/hep/www?irn=#1}{#2}}
\newbox\pippobox

\def\II{\relax{\rm I\kern-.18em I}}

\def\m{\mu}
\def\n{\nu}

\def\t{\theta}
\def\sp{\;\;\;,\;\;\;}

\def\a{\alpha}

\def\tr{\ensuremath{\mathrm{Tr}}}

\title{Chiral symmetry breaking as \\
open string tachyon condensation}

\author{Roberto Casero$^a$, Elias Kiritsis$^{a,b}$,
\'Angel Paredes$^a$\\
~\\
$^a$ CPHT, \'Ecole Polytechnique, UMR du CNRS 7644,
 91128 Palaiseau, FRANCE\\
~\\
$^b$ Department of Physics, University of Crete,
71003 Heraklion, GREECE}

\preprint{\hepth{0702155} \\ CPHT-RR009.0207 }

\abstract{We consider a general framework to study holographically the
dynamics of fundamental quarks in a confining gauge theory. Flavors are
introduced by placing a set of (coincident) branes and antibranes on a
background dual to a confining color theory.
The spectrum contains an open string tachyon and its condensation
describes the $U(N_f)_L \times U(N_f)_R \to U(N_f)_V$ symmetry breaking.
By studying worldvolume gauge transformations of the flavor brane action, we
obtain the QCD global anomalies and an IR condition that allows to fix
the quark condensate in terms of the quark mass. We find the expected $N_f^2$
Goldstone bosons (for $m_q=0$), the Gell-Mann-Oakes-Renner relation
(for $m_q$ small) and the $\eta'$ mass.
Remarkably, the linear confinement behavior for the
 masses of highly excited spin-1 mesons, $m_n^2\sim n$ is
 naturally reproduced.
}


\begin{document}


\section{Introduction}

The original AdS/CFT conjecture
\cite{Maldacena:1997re,WAdS} states the
existence of a stringy description of a conformal, highly symmetric,
quantum field theory. The low energy limit on the string side (supergravity)
corresponds to the strong coupling regime of the gauge theory.
A natural question that has attracted a lot of attention is whether such
a duality can be generalized in order to describe the strong coupling
regime of more realistic gauge theories.
In the last few years we have learned that the
ideas of holography can indeed be applied  to a certain extent in that direction.
In particular, higher-dimensional
duals of (non-supersymmetric)
QCD-like theories have been built and many
qualitative and, surprisingly, some
quantitative features of real-world strong interactions can
be described in such frameworks with
reasonable accuracy.

One of the most important strong-coupling phenomena of QCD is spontaneous
chiral symmetry breaking $U(N_f)_L \times U(N_f)_R \to U(N_f)_V$
by the formation of a non-perturbative quark condensate $\langle \bar q q\rangle \neq 0$
($N_f$ denotes the number of quark flavors).
The goal of this paper is the study of this symmetry breaking using holographic techniques.

We start by briefly reviewing the results obtained on this topic in the existing literature.

This
question was initially addressed in \cite{Babington:2003vm,Kruczenski:2003uq}
(see also \cite{Evans:2004ia}),  by considering
stacks of $N_f$ flavor branes in
non-supersymmetric backgrounds created by $N_c$ color branes.
The chiral symmetry
is a  $U(1)_A$ isometry of the geometry and is spontaneously broken
due to the embedding of the flavor branes. Typically, these
 models are  asymptotically supersymmetric in the UV.
This allows to have good control on quantities like the
quark condensate and the quark mass. Imposing regularity
of the brane configuration, one obtains an IR condition which fixes
the quark condensate in terms of the quark mass, as expected in QCD.
Also, by introducing a small quark mass,
the Gell-Mann, Oakes, Renner (GOR) relation for the pion masses
$$m_\pi^2=-2\frac{m_q \langle \bar q q\rangle }{f_\pi^2}\sp m_q\to 0$$
can be obtained. The ${\cal O}(\frac{N_f}{N_c})$
 mass of the $\eta'$ in this kind of setup was studied
in \cite{Barbon:2004dq}.
The limitation of this approach is that, even for massless quarks,  only the abelian chiral symmetry
is present,
while the non-abelian chiral subgroup of the flavor symmetry
is absent from the beginning,
and its breaking cannot therefore be seen.
{}From
the field theory side, this is due to the existence of a
$\bar q \Phi q$ Yukawa coupling (color and flavor branes are codimension four),
where $\Phi$ is one of the (massive) scalars living on the brane
world-volume~\cite{Kruczenski:2003uq}.

A different approach was followed in \cite{Sakai:2004cn,SS2}. By considering
$N_f$ D8-$\overline {\rm D8}$ pairs in a non-supersymmetric background
corresponding to $N_c$ D4 branes, a $U(N_f)\times U(N_f)$ global symmetry
was introduced. Contrary to the case described earlier,
the chiral symmetry is now realized on the
flavor brane world-volume rather than in the geometry.
The breaking of the full non-abelian chiral symmetry
is due to a
smooth recombination of branes and antibranes in the IR.
By computing the meson spectrum, $N_f^2$
massless Goldstone bosons are found.
One of them, the $\eta'$, is  massless only when $N_c \to \infty$
 and the authors
addressed the problem of finding its ${\cal O}(\frac{N_f}{N_c})$
mass. The chiral anomaly and associated WZW term are nicely reproduced.
Remarkably, quantities (some meson masses, decay constants and couplings)
computed in this holographic setup match reasonably
well the experimental values.
Moreover, the model also incorporates a very natural picture of
chiral symmetry restoration at high
temperature: once a horizon
is formed in the geometry, the branes and antibranes
can fall into it and they do not need to recombine \cite{Aharony:2006da}.

This construction however presents some shortcomings too.
First, there is no
parameter which can be associated to the quark bare masses. By modifying
the flavor brane embedding, one can vary the masses
of the massive mesons,
which can be thought of
as a modification of the constituent quark masses.
However, the pions remain massless, implying that  the bare quark
masses are always vanishing (see
\cite{Aharony:2006da,Casero:2005se,Antonyan:2006vw,Peeters:2006iu}).
This is clearly not a physical feature of QCD. Related to this problem, there
is no parameter that can be identified with the quark condensate,
even though this is a crucial quantity to describe spontaneous
chiral symmetry breaking in QCD.
Another unphysical feature of the model is the absence
of a tower of massive mesons with the quantum numbers of the pion.

Finally, there is an interesting alternative approach,
that has been named AdS/QCD \cite{Erlich:2005qh,DaRold:2005zs}.
It is a bottom-up construction in the
sense that it does not emerge directly  from some known string theory.
Rather, it just assumes the existence of a dual description of QCD living in a five-dimensional,
asymptotically AdS space. The five-dimensional bulk fields
needed to dualize the quark bilinears are introduced by hand, and
spontaneous
chiral symmetry breaking is modeled by giving a vev  to
a bulk scalar by hand.
The model incorporates the existence of Goldstone
bosons, vector meson dominance \cite{DaRold:2005zs} and the
GOR relation for the pion masses \cite{Erlich:2005qh}.
In its simplest form, the five-dimensional
space is taken to be just $AdS_5$ with a hard IR cutoff. However,
modifications of the background have been considered in order to
incorporate the effect in the geometry of the possible
 condensates~\cite{Shock:2006qy} or linear confinement~\cite{Karch:2006pv}.
A drawback of the model is that, since the condensate is 
not determined from some dynamical computation, there is an
extra parameter compared to QCD. Also, being {\it ad hoc},
the models are less satisfying and it would be nice to understand how they
are related to some (possibly non-critical)
string theory which may help fix the background geometry
or the potential for the bulk scalar.
Some progress along this line seems imminent, \cite{UmutElias}.

\vskip .1cm

The goal of this paper is to improve the
present holographic description of chiral
symmetry breaking by proposing a general string theory  construction
which accounts for all relevant phenomena in a natural
and simple way. Flavors are introduced via a stack of $N_f$
overlapping  brane-antibrane pairs. At the same time, we allow
for a non-trivial profile for the open string tachyon field.
This is generally the lightest string mode extending between
branes and antibranes, and it transforms in the bifundamental
representation of the $U(N_f)_L \times U(N_f)_R$ flavor symmetry
group supported by the brane-antibrane pairs. This complex scalar
is therefore the natural candidate to describe chiral
symmetry breaking, as suggested in
\cite{Sakai:2004cn} (see also \cite{Antonyan:2006vw,Sugimoto:2004mh,BCCKP}).
A review on open string tachyon physics can be found in \cite{Sen:2004nf}.

We will study the world-volume effective action for the
brane-antibrane system, including the
tachyon both in the Dirac-Born-Infeld and Wess-Zumino terms.
To keep the focus on the relevant dynamics, we will consider
a simple setup in which only the tachyon and the world-volume gauge
fields of the branes and antibranes are dynamical.
This is the minimal
setup to include the quark bilinears in the dual theory (up to
spin one) and it matches the bulk field
content of the AdS/QCD models \cite{Erlich:2005qh,DaRold:2005zs}.
We will also not worry about the closed string physics. We will keep
the discussion very general, without applying it to any particular
model: we will only need to assume that the background metric is
such that the color theory is confining  (using a general
prescription in \cite{Kinar:1998vq}).

In a few places we will also perform some more explicit computations;
in these cases we will assume that the space is
asymptotically AdS in the~UV.  This will allow us to
apply the general formalism developed in this paper to a relevant and
interesting example, while hopefully clarifying our presentation.

\subsection{Summary of results}

The open string tachyon we introduce transforms in the bifundamental representation of the flavor
symmetry group $U(N_f)_L\times U(N_f)_R$, and couples on the boundary to the quark scalar and pseudoscalar bilinears. Up to a normalization we will introduce later on, we have in fact:
\be
T\leftrightarrow \bar{q}\frac{1+\gamma^5}{2} q\qquad\qquad T^\dagger \leftrightarrow \bar{q}\frac{1-\gamma^5}{2} q
\ee

By considering  brane-antibrane configurations, and including  the
dynamics of the complex open
 string tachyon  scalar $T$, we have been able to provide a general
framework to describe holographically the chiral dynamics of QCD and related theories,
and to reproduce many features of QCD:
\begin{itemize}

\item The UV non-normalizable component
of the tachyon corresponds
to the quark's bare (complex) mass matrix $m_q$.
For simplicity, in this paper we will consider all quarks to have the same (possibly vanishing) mass $m_q$.

\item The UV normalizable mode of the tachyon corresponds, in turn, to turning on an expectation value
for the quark condensate.

\item We show that in a confining background, even for $m_q=0$, the tachyon profile cannot be trivial,
which results in the chiral part $U(N_f)_A$ of the flavor symmetry being always broken, either explicitly or spontaneously.
The construction of this paper, therefore, provides a holographic version of the Coleman-Witten theorem \cite{CW}.

\item The fact that
the tachyon diverges  in the IR (signalling the IR fusion of branes and antibranes) constrains the way it vanishes in the UV:
this allows  to
fix  the value of the quark condensate
$\langle \bar qq\rangle$ in terms of the quark bare mass $m_q$.

\item
We can derive formulae for the anomalous divergences of
flavor currents, when they are coupled to an external source.

\item The WZ part of the flavor brane action gives  the Adler-Bell-Jackiw $U(1)_A$ axial anomaly
\cite{Adler:1969gk}, and an associated Stuckelberg mechanism gives an $O\left(\frac{N_f}{N_c}\right)$  mass
to the would-be Goldstone boson $\eta'$, in accordance with the Veneziano-Witten formula.

\item When $m_q=0$ (i.e. the tachyon has no UV non-normalizable mode),
we find that (in the $N_c\rightarrow \infty$ limit) the meson
spectrum contains $N_f^2$ massless pseudoscalars, the
$U(N_f)_A$ Goldstone bosons.

\item Studying the spectrum of highly excited spin-1 mesons, we find the expected property of linear confinement: $m_n^2\sim n$.

\item When we consider an asymptotically AdS space in the UV, we can precisely reproduce the
GOR relation for the mass of the pions in the case of a small but non-zero quark bare mass.

\end{itemize}

On the other hand, the main limitations
our approach still presents  are the following:

\begin{itemize}

\item The details of the tachyon potential that one should insert in the DBI action
are not rigorously known. Despite this, its gross features suffice for extracting qualitative information.

\item A background in which the brane-antibranes are on top of each
other (and with no Wilson line turned on), would probably require large
curvature, as we  explain. Then, a reliable background metric cannot be
found by solving Einstein's equations.

\item We find that the slope of the
linear confinement relation  is not the same for vector and axial mesons.
This may be due to the fact that
it is not clear whether the DBI action we
use to include the effects of the dynamics of the
open string tachyon remains valid in the region
where the tachyon and its derivative diverge.

\end{itemize}

However we found  encouraging that a large  amount of expected QCD features  can be extracted
from our construction despite the  misgivings above.

\section{The general construction}
\label{sect: general}

The general setup we consider consists of a system of $N_f$ overlapping
${\rm D}p$-$\overline {{\rm D}p}$ flavor brane-antibrane pairs in the background
associated to $N_c$ D$q$ color branes. In critical, ten-dimensional
string theory, we will need to impose the condition $p+q=12$ and require the color
and flavor branes to be codimension 6, in order to get
the correct WZ couplings.\footnote{The reason why is very simple
to explain. In section \ref{sect: anomaly} we will be looking for
a coupling on the flavor D$p-\overline{{\rm D}p}$ branes which involves
6-dimensional gauge field forms. This can only come from the
Wess-Zumino coupling of the $p$-branes to a magnetic  $k$-form
potential, where $k=p+1-6=p-5$. Now D$q$-branes source
a $C_{q+1}$ RR potential, which is parallel to the
source branes and generically depends on the radial
direction alone. Imposing 4-dimensional Lorentz invariance
of the background ensures that the Hodge dual of $C_{q+1}$
is a magnetic $\tilde{k}$-form, with $\tilde{k}=10-(q+2)-1=7-q$.
Imposing $k=\tilde{k}$ finally gives the condition $p+q=12$.}
For instance, one can think of a
${\rm D}3$-${\rm D}9$-$\overline {{\rm D}9}$ intersection as
in~\cite{Sugimoto:2004mh} or a ${\rm D}4$-${\rm D}8$-$\overline {{\rm D}8}$
intersection
similar to the Sakai-Sugimoto model \cite{Sakai:2004cn}.
Notice, though, that the general formalism we develop can also be applied to lower
dimensional  setups
provided the WZ couplings are the
obvious adaptation of those of section \ref{sect: WZ}.
Interesting examples of this kind are the AdS/QCD models
or non-critical string  constructions
(${\cal N}=1$ theories
with space-time filling
brane-antibrane flavors have been discussed in
\cite{Klebanov:2004ya,BCCKP,cigar} and ${\cal N}=0$
in~\cite{BCCKP,Casero:2005se}).

In the spectrum of stacks of overlapping brane-antibrane
pairs there is a tachyonic, complex scalar, open string mode,
which transforms in the bifundamental representation of the $U(N_f)_L \times U(N_f)_R$ flavor
 symmetry group supported by the brane-antibrane pairs. As we argued in the introduction,
  we can in general associate chiral symmetry breaking to a vacuum expectation value for this field.
Near the UV region (in our coordinates this is around $z=0$), the tachyon's vev vanishes and the full
$U(N_f)_L \times U(N_f)_R$ is present, as in the lagrangian
of massless QCD. As we approach the IR, we prove in section \ref{sect: anomaly} that if the
theory is confining the tachyon must acquire a non-trivial vev, breaking the symmetry to its diagonal $U(N_f)_V$ subgroup, (see also
section \ref{sect: vev}).
Concretely, we show that the vev reaches
infinity at a finite distance in the bulk, a point that we denote
by $z_{IR}$. In a sense, this process can be thought of as brane-antibrane
recombination. In the minimal construction that we consider in the rest of this paper,
$z_{IR}$ coincides with the end of space.
Figure \ref{scheme} presents a general profile for the tachyon in the setup we just described.
\FIGURE[!h]{\epsfig{file=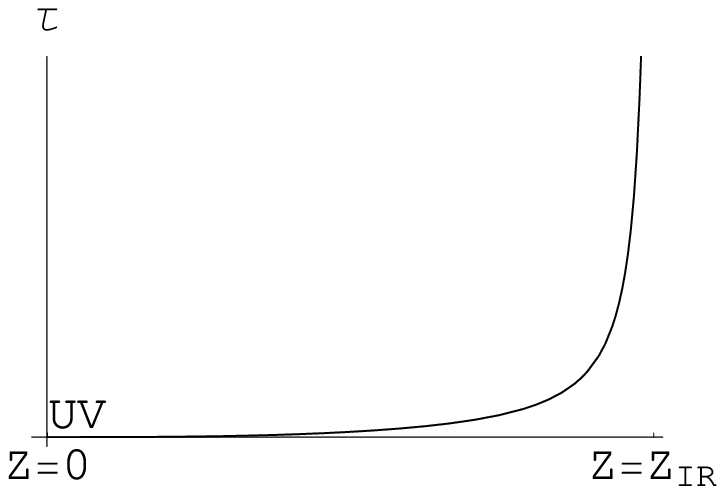,width=0.6\textwidth}
\caption{A scheme of the setup. Generically, the tachyon vanishes in
the UV where one has a stack of brane-antibrane pairs. In the
IR, it diverges and the brane-antibrane pairs end smoothly. Here $\tau$ stands for the modulus of the tachyon field.}
\label{scheme}}

For the present discussion, it will be very important to consider
the completion of the world-volume action of the brane-antibrane
system to include also the couplings to the open string tachyon.
As usual, this action is composed  of a Dirac-Born-Infeld and a
Wess-Zumino term.
In the rest of this section and in section \ref{sect: DBI},
we study the physics of the DBI piece, which yields
the vacuum of the configuration and the meson mass spectrum. The WZ part
will be studied in section \ref{sect: WZ}. {}From it, we derive
the holographic description of the parity and charge conjugation symmetries,
the global anomalies, and a holographic
 Coleman-Witten theorem.

\subsection{The degrees of freedom: a minimal setup}
\label{sect: dof}

As  proposed in \cite{Sen:2003tm, Garousi:2004rd}, the semiclassical
action for a D$p$-$\overline{{\rm D} p}$ system which
includes the physics of the open string tachyon  is a
generalization of the usual Dirac-Born-Infeld  action for D-branes.

We start from the simpler configuration of a
single brane-antibrane pair. In this case the proposed DBI action reads:
\be
S=-\int d^{p+1}x\, e^{-\phi}\, V(\tau^2,Y^I_L-Y^I_R,x)
\left(\sqrt{-\det {\bf A}_L}+\sqrt{-\det {\bf A}_R}\right)
\label{generalact}
\ee
where (we will set $2\pi\a'=1$ from now on, and only reinsert
$\a'$  when needed):
\bear
&&A_{(i)MN}=P[g+B]_{MN} + F^{(i)}_{MN} + \partial_M Y_{(i)}^I
\partial_N Y_{(i)}^I+\frac{1}{\pi}(D_M T)^* (D_N T)+
\frac{1}{\pi}(D_N T)^* (D_M T)\rc
&&F_{MN}^{(i)} = \partial_M A_N^{(i)} -\partial_N A_M^{(i)}\,,\qquad
D_M T = (\partial_M + i A_M^L- i A_M^R)T
\label{defs1}
\eear
and $V(\tau^2,Y^I_L-Y^I_R,x)$ is the tachyon potential.
The complex tachyon is denoted by $T \equiv \tau e^{i \theta}$, the indices
$i=L,R$ denote the brane or antibrane, the $Y^I_{(i)}$ are the transverse
scalars, and $A^{(i)}$ is the world-volume gauge field. Here and in the following, capital latin characters, $M,N$, denote
 world-volume directions, which include in particular the
 Minkowski directions, which we will denote by greek letters $\m,\n$, and the
 holographic radius which will be referred to as $z$.

For simplicity, we will not consider a background $B$-field. We will
also ignore the transverse scalars living on the flavor branes.
These modes, which are generally present in a critical string setup,
do not have any obvious QCD interpretation and would not play any
relevant role in the present discussion.
Thus, we only consider the degrees of freedom coming from the
open string tachyon and the
world-volume gauge fields along the Minkowski or
the holographic directions $A^{(i)}_\m, A^{(i)}_z$.
This field content is enough to include the basic QCD quark bilinears
up to spin 1 and coincides with the one of the AdS/QCD models.

Higher spin operators correspond to stringy excitations
and cannot be addressed in this formalism. They could be introduced
by hand as in \cite{Katz:2005ir}.

Since we are ignoring transverse scalars, for the tachyon potential we take the expression for overlapping brane-antibranes\footnote{If one considered more complicated
setups with non-trivial brane profiles, the potential
should depend also on the brane-antibrane distance.
For a discussion in a related setup, see \cite{Antonyan:2006vw}.} derived in boundary string field
theory~\cite{Kutasov:2000aq}, \cite{Minahan:2000tf}\footnote{
Unfortunately,
there are not unified conventions
in the literature about the definition of
$T$. The unambiguous statement is that the mass squared
of the open string
tachyon in flat space is $m_T^2=-\frac{1}{2\a'}$.
Equation (\ref{gaussian}) is thus consistent with
(\ref{generalact}) and (\ref{defs1}) since, then, for small $\tau$ the
action is proportional to
$S\propto \int \left(\frac12 (\partial \tau)^2 - \frac12 \frac{1}{2\alpha'}
\tau^2\right)$, where we have considered flat space and reinserted
$\a'$.}:
\be
V(\tau^2)=T_p\, e^{-\tau^2}
\label{gaussian}
\ee
One should keep in mind that this expression is not rigorously justified
since there can be non-trivial field redefinitions between the two
formalisms. Moreover the boundary string field theory computation is performed in flat space
and it is not trivial at all that the result should apply  unchanged to curved backgrounds.
Remarkably, in section
\ref{sect: linear} we will find the nice feature of linear confinement
using this gaussian behavior for the potential.
Most of the rest of the discussion is insensitive to the
precise expression of the tachyon potential.

After presenting the simpler abelian case, we can now go  to considering a stack of $N_f> 1$ brane-antibrane pairs,
 which is what we are eventually interested in.  The natural
non-abelian generalization of  (\ref{generalact}) is:
\be
S=-\int d^{p+1}x \
{\rm SymTr}\left[ e^{-\phi} V(TT^\dagger)
(\sqrt{-\det {\bf A}_L}+\sqrt{-\det {\bf A}_R})
\right]
\label{generalact2}
\ee
where SymTr is a symmetric trace as defined in \cite{Tseytlin:1997cs}.
The analogous proposal for the action of a stack of non-BPS
D-branes was made in \cite{Garousi:2000tr} and can be related
to systems of branes-antibranes along the lines of \cite{Garousi:2004rd}.
Although the symmetric trace prescription for the non-abelian action is  known to be corrected at higher
orders in $\a'$, this will not be relevant for our discussion.
All that is important is that the potential appears in a single trace, and this is guaranteed by the
large $N_c$ limit, \cite{CW}.

The gauge
group supported by the brane-antibrane system
is $U(N_f)_L \times U(N_f)_R$. The gauge fields $A_{L,R}$
transform in the adjoint representation of $U(N_f)_L$ and
$U(N_f)_R$ respectively, while the tachyon $T$ transforms in the
bifundamental of $U(N_f)_L \times U(N_f)_R$, (that is,
 in the antifundamental of $U(N_f)_L$ and in the
fundamental  of $U(N_f)_R$). Its hermitian
conjugate $T^\dagger$ transforms in the opposite way.
For more details on our choice of conventions,
we refer the reader to appendix \ref{app: conventions}.

\subsection{The vev for the tachyon field}
\label{sect: vev}

In this section we want to determine the vacuum configuration for
a stack of brane-antibranes over a confining vacuum. It is trivial
to show that the equations of motion allow for the gauge fields to
be consistently set to zero, $\langle A_L\rangle = \langle A_R\rangle=0$. As we mentioned in the
introduction, and we will explicitly show in section
\ref{sect: anomaly}, anomaly considerations regarding the
WZ part of the action of a stack of brane-antibranes over a
confining vacuum require that  $\langle\tau\rangle |_{z_{IR}}\to\infty$.
This necessitates  that we allow for a non-trivial profile of the
tachyon in the vacuum configuration.

Four-dimensional Poincar\'e invariance determines the
general form of the bulk five-dimensional metric. In the mostly
plus signature this reads:
\be\label{gen 5d met}
ds_5^2 = g_{xx}(z) \eta_{\m\n} dx^\m dx^\n + g_{zz}(z) dz^2
\ee

The non-abelian action (\ref{generalact2}) is a very complicated object, where different components
of the fields mix because of  non-trivial commutation relations.  We start from the simpler
 $N_f=1$ case, where these complications do not arise. {}From (\ref{generalact}), it is
consistent to set the tachyon phase to a constant.

We are thus left with the world-volume
action for the modulus of the tachyon $\tau$, which,
upon reducing to five dimensions reads:
\be
S= - 2 \int d^4x\, dz\, e^{-\phi_{eff}}\, V(\tau^2)\, g_{xx}^2
\sqrt{g_{zz} + \frac{2}{\pi}(\partial_z \tau)^2}\;.
\label{tachyonac}
\ee
The prefactor
$e^{-\phi_{eff}}(z)$ is defined as the dilaton exponential
times whatever comes from integrating the transverse,
spectator, $p-4$ dimensions. Schematically:
\be
\begin{split}
&e^{-\phi_{eff}}(z)\, V(\tau^2(z)) \, g_{xx}^2
\sqrt{g_{zz} + (\partial_z \tau)^2}\equiv\\
&\equiv \int_{\mathcal{N}^{p-3}} d^{p-4}y \,e^{-\phi}\,V(\tau^2,Y_L^I-Y_R^I,x) \left(\sqrt{-\det {\bf A}_L}+\sqrt{-\det {\bf A}_R}\right)
\end{split}
\ee
where $\mathcal{N}^{p-3}$ is the internal part of the flavor branes world-volume.
\footnote{The reader might be puzzled by the  approximation involved in
discarding all dependence on the internal world-volume scalars. It should be noted, though,
that our ansatz for the five-dimensional metric (\ref{gen 5d met}) is completely general, being
 constrained only by four-dimensional Poincar\'e invariance. The vacuum configuration of the
 internal scalars will only affect the explicit form of the metric coefficients and the radial
 dependence of the tachyon modulus and dilaton. Since our arguments are completely independent
  of the particular form of the confining, reduced, five-dimensional  background, all the following
  results apply regardless what the critical string setup is.}

 The Euler-Lagrange equation for
$\tau$ obtained from (\ref{tachyonac}) reads:
\be
\partial^2_z \tau +\frac{2}{\pi}
\frac{\partial_z ( g_{xx}^2 e^{-\phi_{eff}})}{
g_{zz}g_{xx}^2 e^{-\phi_{eff}}}(\partial_z \tau)^3
+ \left(  \frac{\partial_z ( g_{xx}^2 e^{-\phi_{eff}})}{
g_{xx}^2 e^{-\phi_{eff}}} - \frac{\partial_z g_{zz}}{2g_{zz}}  \right)
\partial_z \tau
+ 2\tau
\left(\frac{\pi}{2} g_{zz} +(\partial_z \tau)^2 \right)=0
\label{explicittaueq}
\ee
This is a second order differential equation which therefore
has two integration constants.
Since $\tau$ is dual to the quark bilinear, we know on general
AdS/CFT grounds that, by looking at the UV behavior of $\tau$, these
two constants can be related to  the quark bare mass and the quark
condensate. We now assume that the space is asymptotically AdS, that is
$g_{xx}(z)\simeq g_{zz}(z)\simeq R_{AdS}^2/z^2+\ldots$ and
$e^{-\phi_{eff}}\simeq const\equiv e^{-\phi_0}$
for small $z$.
The tachyon field is dual to the
quark bilinear which has dimension 3.
Thus, the usual relation between the mass of a
five-dimensional scalar field and the conformal
dimension of its dual operator, $\Delta (\Delta - 4)=m^2 R_{AdS}^2$,
fixes $m_T^2 R_{AdS}^2 =-3$, which in turn implies that, if $m_T^2=-\frac{1}{2\alpha'}=-\pi$, the radius of $AdS$ should be small:  $R_{AdS}^2=6\alpha'=\frac{\pi}{3}$.
This leads to the following
UV behavior (in order to simplify notation the vacuum expectation
value $\langle\tau\rangle$ will be  denoted
 just as $\tau$ from now on):
\be
R_{AdS}^\frac32 \tau_{can} \equiv
R_{AdS}^\frac32 \left(-\frac{4 T_p}{m_T^2 e^{\phi_0}}\right)^\frac12
\tau = m_q (z + \dots) +  \sigma (z^3 + \dots)
   \qquad \textrm {(small $z$)}\,.
\label{UVtau}
\ee
We have written this expression in terms of
the canonically normalized $\tau_{can}$, such that the
boundary coupling is
$R_{AdS}^\frac32 \int d^4x  \,\tau_{can}\,z^{-1} \bar q q$.
The parameter $\sigma$ can be related to the quark
condensate $\langle\bar q q\rangle$, see appendix
\ref{holorenorm} for details.
The UV behavior (\ref{UVtau}) is the same
 as the one for the scalar breaking the chiral symmetry
in the AdS/QCD models \cite{Erlich:2005qh,DaRold:2005zs}.

We now want to prove that the IR condition $\tau|_{z_{IR}}
\to \infty$ gives a condition on $\tau(z)$, fixing $\sigma$, and therefore
the condensate,
in terms of the mass $m_q$.

A general result of the gauge-gravity correspondence states  that a
sufficient condition for confinement is that at some $z_{div}$,
$g_{zz}(z_{div})\to \infty $  while $g_{xx}(z_{div})\neq 0 $ and
$\partial_z g_{xx}(z_{div})<0$ \cite{Kinar:1998vq}. We will assume this is the case
 for our background and in particular, after possibly redefining $z$, we fix the divergence to be a single pole
 $g_{zz} = b (z_{div} - z)^{-1}$ near $z=z_{div}$.

 This is the behavior, for instance, in Witten's dual of
 Yang-Mills \cite{Witten:1998zw}. Inspecting
 the leading terms for large $\tau$ of (\ref{explicittaueq}),
 we see that the tachyon can only diverge at $z_{IR}=z_{div}$ and there it
 diverges as $\tau \propto (z_{IR} - z)^{-a}$, where
 $a=-\pi b\left[\frac{g_{xx}^2 e^{-\phi_{eff}}}
 {\partial_z (g_{xx}^2 e^{-\phi_{eff}})}\right]_{z=z_{IR}}$
 is a positive number since the derivative inside the bracket
 is typically negative.

 We warn the reader that this result should be taken with a grain of salt because
 it relies on the behavior of (\ref{explicittaueq}) when
 $\tau, \dot\tau \to \infty$,
 which lies outside the regime of validity of the DBI action
(\ref{tachyonac}). Nonetheless, the general lesson one should learn from this
computation is that the $\tau|_{z_{IR}} \to \infty$
consistency condition
constrains the IR behavior of the tachyon. Physically, this fixes
 the quark condensate in terms of the quark
mass\footnote{It is natural that for a fixed $m_q$, the conditions
stated above fix uniquely $\sigma$. However, within the present
very general formalism, this uniqueness cannot be proved. We
will come back to this in section \ref{sect: CW}},
in the spirit of~\cite{Babington:2003vm,Kruczenski:2003uq}.

At this point we can consider the more general non-abelian case $N_f>1$.
Throughout the paper, we will consider for simplicity
that all $N_f$ flavor fields have  the same mass $m_q$.
The symmetrized trace in the action (\ref{generalact2}) might seem
to make it a lot more complicated to find a vacuum configuration for
the tachyon matrix $T$. Fortunately
we can show that there is a perturbatively stable configuration
 given by $N_f$ identical copies of the solution
for the abelian case we described above
\be
\langle T\rangle = \tau(z)\ \II
\label{vacuum}
\ee
In fact, we can apply here a result that
was derived in \cite{Kruczenski:2003uq} for a different setup.
The reasoning in \cite{Kruczenski:2003uq}
showing that the vacuum configuration where vevs are proportional
to identity matrices, like (\ref{vacuum}) in our case,  is a stable
(up to quadratic order) solution of the non-abelian DBI action
is very general. It only relies on having a single trace
in front of the action and the fact that
the difference between abelian and non-abelian DBI actions
consists entirely of terms involving commutators.
The argument does not prove that this is the global minimum.
There could in principle be a different configuration with lower
energy, but this is an unlikely possibility and we will assume
that (\ref{vacuum}) describes the vacuum of the theory.

Once we have established that $\langle T\rangle $ is proportional to the
identity matrix, the equation of motion for
$\tau(z)$ is the same as (\ref{explicittaueq}) as well as the UV and
IR conditions, since for the configuration (\ref{vacuum}) and up
to the order we need here, the non-abelian action (\ref{generalact2})
is actually the sum of $N_f$ identical copies of the abelian
action (\ref{generalact}).

Looking at how the fields transform under flavor symmetry transformations (\ref{gaugevar}), it is clear that the vacuum (\ref{vacuum})
spontaneously\footnote{From the field theory point of view the nature of the breaking depends on
 the asymptotic UV behavior of the tachyon. When $m_q\neq 0$, the vacuum configuration of the
  tachyon is not normalizable and we are adding a mass term for the quarks to the lagrangian:
   the breaking is explicit. If instead $m_q= 0$,  the tachyon is normalizable, and we are turning on
an expectation value for the quark bilinear in the original theory, which makes the symmetry breaking
spontaneous. From the gravity point of view, instead, both for vanishing or non-zero $m_q$, the
 breaking is spontaneous  since it is determined by a non-trivial expectation value.} breaks
$U(N_f)_L \times U(N_f)_R$ into its diagonal subgroup $U(N_f)_V$,
where the vectorial
gauge transformation is $V_L = V_R$.

When $m_q=0$, this spontaneous breaking
implies the existence of $N_f^2$ Goldstone bosons. We will
explicitly find them in the spectrum in section \ref{sect: Goldstone}
(see also section \ref{sect: 3point} for an indirect argument).
We conclude this section by noting that the fact that the symmetry is broken down to $U(N_f)_V$, and not
further, is also related
to a theorem by Vafa and Witten \cite{VafaWitten} that states
that vectorial symmetries
cannot be spontaneously broken.

\section{The D-brane Wess-Zumino sector}
\label{sect: WZ}
\setcounter{equation}{0}

We now consider the WZ coupling of the $N_f$
${\rm D}p\,$-$\overline{{\rm D}p}$ branes to the
background RR fields, including the
tachyon dependence.
This part of the action  is given by (the following expression was
proposed in \cite{tach in CS}
and proved in \cite{KrausLarsen,TTU} using boundary string field theory):
\be\label{WZtach}
S_{WZ}=T_{p} \int_{\Sigma_{p+1}} C\wedge \mathrm{Str} ~\exp\left[{i
2\pi\a'\mathcal{F}}\right]
\ee
where $\Sigma_{p+1}$ is the world-volume of the
${\rm D}p\,$-$\overline{{\rm D}p}$ branes,
$C$ is a formal sum of the RR potentials $C=\sum_n (-i)^{\frac{p-n+1}{2}}C_n$,
and $\mathcal{F}$ is the curvature of a superconnection
${\cal A}$.
In terms of the tachyon field matrix $T$
and the gauge fields  $A_L$ and  $A_R$
living respectively on the branes and
antibranes, they are
(from now on, we will set $2\pi \alpha'=1$
again and use the notation of
\cite{KrausLarsen}):
\be
i\mathcal{A}=\left(\begin{array}{cc} iA_L & T^\dagger\\
T & iA_R\end{array}\right)\,,\qquad
i\mathcal{F}=\left(\begin{array}{cc} iF_L-T^\dagger T & DT^\dagger\\
DT & iF_R-TT^\dagger\end{array}\right)
\label{AFdef}
\ee
In appendix \ref{superappendix} we review the relevant definitions
and properties of this {\it supermatrix}
formalism (in particular, notice the definition of the
supermatrix product).

The superconnection is defined as:
\be
{\cal F} = d{\cal A} - i {\cal A} \wedge {\cal A}
\ee
and satisfies the Bianchi identity:
\be
d{\cal F} - i {\cal A} \wedge {\cal F} + i {\cal F} \wedge {\cal A} = 0
\ee
Using this identity and the cyclic property of the supertrace
(\ref{cyclic}) it is immediate to check that
$\mathrm{Str} ~e^{i\mathcal{F}}$ is a closed form and  therefore that
(\ref{WZtach}) is invariant under RR gauge transformations.
At least locally, there exists then a form $\Omega$ such that:
\be
d\Omega=\mathrm{Str} \,e^{i\mathcal{F}}
\ee

The case of interest here is the WZ coupling on the
${\rm D}p\,$-$\overline{{\rm D}p}$ flavor branes in the background
of $N_c$ D$q$-color branes. As explained at the beginning of section
\ref{sect: general}, we have $p+q=12$ and the electric RR potential sourced by the color branes is then
 $C_{q+1}=C_{13-p}$, while its magnetic dual is $C_{p-5}$.
Formula (\ref{WZtach}) reduces to:
\be\label{WZtach step1}
S_{WZ}=\, i\,T_{p} \int_{\Sigma_{p+1}} C_{p-5}\wedge\left.
\mathrm{Str}\ e^{i\mathcal{F}}\right|_{6\mathrm{-form}}
=i(-)^{p} \, T_{p} \int_{\Sigma_{p+1}} F_{p-4}\wedge
\Omega|_{5\mathrm{-form}}
\ee
When the branes worldvolume $\Sigma_{p+1}$ has boundaries, the two
expressions in (\ref{WZtach step1}) differ by boundary terms, and
are therefore not equivalent. In this case, which is the relevant
one for this paper, it is argued in \cite{Green:1996dd} (for
intersecting branes) that the correct form for the action  is
the last one in (\ref{WZtach step1}).
Since $\Omega|_{5\mathrm{-form}}$
is defined only up to an exact form, there is then a possible
ambiguity in the definition of the WZ action (\ref{WZtach step1}).
We will assume that the correct form for $\Omega|_{5\mathrm{-form}}$
respects the discrete symmetries of
$\mathrm{Str}\,e^{i\mathcal{F}}|_{6\mathrm{-form}}$, which we
 will discuss in the following subsection \ref{sect: PC}.

Notice that the potential
sourced by the D$q$-branes has indices parallel to the branes and
generically depends on the radial direction.
Moreover, 4-dimensional Lorentz invariance  of the background
(plus possibly a radial redefinition) prevents the metric from
having non-zero off-diagonal terms in the directions $(0,1,2,3,z)$.
These two facts ensure that $F_{p-4}$ does not have legs along
the $(0,1,2,3,z)$ directions.
Thus, the only components
of the five-form $\Omega|_{5\mathrm{-form}}\equiv\Omega_5$ which we are
interested in  are those along $x^0$, $x^1$, $x^2$, $x^3$ and $z$
(we will call ${\mathcal{M}_5}$ this five-dimensional space).
Similarly to section \ref{sect: general}, we assume at this
point that $\Omega_5$ does not depend on the directions
along which $F_{p-4}$ lies, as expected on physical grounds.
Then we can
 integrate out those directions
in (\ref{WZtach step1}) to obtain the 5-dimensional effective
Chern-Simons action:
\be\label{5dCS}
S_{CS}= i(-)^{p}\, T_{p} \int_{\Sigma_{p+1}\cap\mathcal{M}_5^\bot}
F_{p-4} \cdot
\int_{\mathcal{M}_5} \Omega_5 =
 (-)^{p} \frac{i N_c}{(2\pi)^5 \a'^3} \int_{\mathcal{M}_5} \Omega_5
\ee
where we have used the quantization condition
$\frac{1}{2\kappa_{(10)}^2}\int F_{p-4}=N_c T_{12-p}$
and the explicit value of the D-brane tensions.

\subsection{Discrete symmetries $(P,C)$}
\label{sect: PC}

The six-form $\mathrm{Str}\,e^{i\mathcal{F}}|_{6\mathrm{-form}}$
appearing in (\ref{WZtach step1})
is invariant under two  discrete symmetries, $P$ and $C$, that we
will present in this subsection. Thus, an
$\Omega_5$ invariant under $P$, $C$ exists.
It follows then from (\ref{5dCS}) that the CS
action is invariant under the same symmetries, which acquire a
fundamental physical interpretation in the dual field theory.

\subsection*{Parity}

The six-form $\mathrm{Str}\,e^{i\mathcal{F}}|_{6\mathrm{-form}}$ is left invariant by the change $P\equiv P_1\cdot P_2$, where:
\be
P_1: A_L \leftrightarrow A_R \,,\quad T\leftrightarrow T^\dagger \qquad \mathrm{and}\qquad P_2:
(x_1,x_2,x_3) \rightarrow (-x_1,-x_2,-x_3)
\label{Ptransf}
\ee
Notice that $P_1$ also transforms  $DT \leftrightarrow DT^\dagger$,
$F_L \leftrightarrow F_R$. Indeed we can easily check that:
\bear
&\mathrm{Str}\ e^{i\mathcal{F}} =\mathrm{Str}
\left[\exp\left( \begin{array}{cc} iF_L-T^\dagger T & DT^\dagger
\\ DT & iF_R-TT^\dagger   \end{array} \right)\right] \xrightarrow{P_1}& \rc[5pt]
&\mathrm{Str}\left[\exp\left( \begin{array}{cc} iF_R-TT^\dagger   & DT
\\ DT^\dagger & iF_L-T^\dagger T\end{array} \right)\right] =
-\mathrm{Str}
\left[\exp\left( \begin{array}{cc} iF_L-T^\dagger T & DT^\dagger
\\ DT & iF_R-TT^\dagger   \end{array} \right)\right]\;.&
\label{chiral transf}
\eear
For the last equality we have just swapped the ordering of the
blocks inside the matrix, picking up a minus sign
due to the definition of the supertrace
(\ref{supertrace}). The overall
minus sign in (\ref{chiral transf}) is compensated by the minus sign
acquired by the form $dx_1 \wedge dx_2 \wedge dx_3$ under the action of $P_2$, and the CS action (\ref{5dCS}) is
 invariant under $P$.

 We conclude that four-dimensional parity is equivalent to
 charge conjugation for the D-branes (interchanging $D\leftrightarrow\bar D$).

\subsection*{Charge conjugation}

The other discrete symmetry of $\mathrm{Str}\,e^{i\mathcal{F}}|_{6\mathrm{-form}}$, and consequently of the CS action (\ref{5dCS}), is:
\be
C:\qquad
A_L \to -A_R^t \,,\quad   A_R \to -A_L^t \,,\quad
T\to T^t\,,\quad
T^\dagger \to \left( T^\dagger\right)^t
\label{Ctrans}
\ee
where the superscript $t$ denotes matrix transposition.
Notice that under (\ref{Ctrans}), the field strengths and tachyon
covariant derivatives transform as
$F_L \to -F_R^t$, $F_R \to -F_L^t$, $DT \to DT^t$,
$DT^\dagger \to \left( DT^\dagger\right)^t$.
The supertrace transforms as:
\small
\bear
&\mathrm{Str}\ e^{i\mathcal{F}} =\mathrm{Str}
\left[\exp\left( \begin{array}{cc} iF_L-T^\dagger T & DT^\dagger
\\ DT & iF_R-TT^\dagger   \end{array} \right)\right] \xrightarrow{C}& \rc[5pt]
&\mathrm{Str}\left[\exp\left( \begin{array}{cc} -iF_R^t -(TT^\dagger )^t  & \left( DT^\dagger\right)^t
\\ DT^t & -iF_L^t-(T^\dagger T)^t \end{array} \right)\right] =
-\mathrm{Str}
\left[\exp\left( \begin{array}{cc}  -iF_L^t-(T^\dagger T)^t  & DT^t
\\  \left( DT^\dagger\right)^t & -iF_R^t -(TT^\dagger )^t   \end{array} \right)\right]=&\rc[5pt]
&-\mathrm{Str}
\left[\exp\left( \begin{array}{cc} -iF_L-T^\dagger T & -i DT^\dagger
\\ -i DT & -iF_R-TT^\dagger   \end{array} \right)^{pt}\right]=
-\mathrm{Str}
\left[\exp\left( \begin{array}{cc} -iF_L-T^\dagger T & -i DT^\dagger
\\ -i DT & -iF_R-TT^\dagger   \end{array} \right)\right]
\label{Ccheck}
\eear
\normalsize
For the  identity in the second line we have
again swapped the ordering of the
blocks picking up a minus sign. In the
equality of the following line we used the definition of
{\it pseudotranspose}~(\ref{defstrans}). Finally in  the last one, we used
property (\ref{propstrans}) and the fact
that the supertrace does not change under
 pseudotransposition.
The matrix of the last expression is not the same as the untransformed one
of the first line of (\ref{Ccheck}). Therefore, (\ref{Ctrans}) does not leave
the supertrace invariant. However, we are only interested in the 6-form
within the full expression of the supertrace.
Noticing that all  two-forms ($F_L, F_R, DT\wedge DT^\dagger$)
inside the matrix of the last line have picked up a minus
sign  compared to the initial expression, we find that the 6-form (in fact, all the $4k+2$-forms)
is left invariant. 

If one wanted to maintain invariant the $4k$-forms of the supertrace (which can be
related to gauge theories in $4k-2$ dimensions in the same way as we relate the 6-form to a
four-dimensional field theory), then the transformation would be $A_L \to -A_L^t \,,\   A_R \to -A_R^t \,,\
T\to \left( T^\dagger\right)^t\,,\
T^\dagger \to T^t$. This is in agreement with the fact that charge conjugation in
$4k-2$ dimensions does not change chirality.

\vskip .2cm

Table 1 summarizes the transformation properties of the
different degrees of freedom.
In particular, notice that the vacuum condensate (\ref{vacuum})
transforms as $0^{++}$ as expected for the $\langle \bar q q\rangle $
bilinear.

\begin{table}[!h]
\label{table1}
\begin{center}
\begin{tabular}{|c|c|c|c|c|}
 \hline
 World-volume field & $T+T^\dagger$
 & $i(T-T^\dagger)$ & $\frac{(A_L+A_R)_\m}{2}$ &
 $\frac{(A_L-A_R)_\m}{2}$ \\ \hline
 $J^{PC}$ & $0^{++}$ & $0^{-+}$ & $1^{--}$ &  $1^{++}$ \\ \hline
\end{tabular}
\caption{\label{t1}A summary of the
world-volume fields, their spins and their transformation properties
under parity and charge conjugation.
In the gauge we will use below (see section
4.1 for more details), $(A_L - A_R)_z$ and
$(A_L + A_R)_z$ are set to zero while the longitudinal component
of $(A_L - A_R)_\mu$ combines with the pseudoscalars.}
\end{center}
\end{table}

\subsection{The chiral anomaly: external currents}
\label{sect: anomaly}

We now want to study the anomaly of the chiral symmetry
when the flavor currents are coupled to external sources,
{\it i.e.} study the WZ term as an action
for the flavor brane world-volume gauge fields ($A_L, A_R$)
which couple in the boundary to combinations
of the vector and axial currents, (see table \ref{t1}).
As is usual for Chern-Simons terms, a gauge transformation
does not leave the action (\ref{5dCS}) invariant
but produces a boundary term.
As in previous cases, \cite{WAdS},
 this term is matched with
the global anomaly of the dual field theory.

We set the tachyon to its vacuum value (\ref{vacuum}).
It is a straightforward although lengthy
computation to expand the supertrace
in (\ref{WZtach step1}).
\be
\begin{split}
&\mathrm{Str}\ e^{i\mathcal{F}}\big|_{6\mathrm{-form}}=\frac{1}{6}\tr\,e^{-\tau^2}\Big\{-i F_L\wedge
F_L\wedge F_L+i F_R\wedge F_R\wedge F_R \;+\\
&+2i\tau d\tau \wedge (A_L-A_R)\wedge \left(F_L\wedge F_L+
\frac{1}{2}F_L\wedge F_R+\frac{1}{2}F_R\wedge F_L +F_R\wedge
F_R\right)+\\[1pt]
&+\tau^2 (A_L-A_R)\wedge  (A_L-A_R)\wedge  (F_L
\wedge F_L-F_R\wedge F_R)+\tau^2 (A_L-A_R)\wedge F_R
\wedge (A_L-A_R) \wedge F_L+\\[7pt]
&-\tau^3 d\tau \wedge (A_L-A_R)\wedge (A_L-A_R) \wedge
(A_L-A_R) \wedge (F_L+F_R)+\\[3pt]
&+\frac{i}{4} \tau^4 \wedge (A_L-A_R)\wedge (A_L-A_R)
\wedge (A_L-A_R) \wedge (A_L-A_R) \wedge (F_L-F_R)\\[3pt]
&-\frac{i}{10}\tau^5 d\tau\wedge (A_L-A_R)\wedge (A_L-A_R)
\wedge (A_L-A_R) \wedge (A_L-A_R) \wedge (A_L-A_R)\Big\}
\end{split}
\label{6form}
\ee
It is now not very hard to find a 5-form $\Omega_5$
such that $d\Omega_5=\mathrm{Str}\ e^{i\mathcal{F}}\big|_{6\mathrm{-form}}$.
 We can write:
\be\label{5-form}
\begin{split}
\Omega_5&=\frac{1}{6}\tr \,e^{-\tau^2}\left\{ -iA_L
\wedge F_L\wedge F_L +\frac{1}{2}A_L\wedge A_L \wedge
A_L \wedge F_L +\frac{i}{10} A_L\wedge A_L \wedge A_L
\wedge A_L\wedge A_L\right.+\\
&+iA_R\wedge F_R\wedge F_R -\frac{1}{2}A_R\wedge A_R
\wedge A_R \wedge F_R -\frac{i}{10} A_R\wedge A_R
\wedge A_R\wedge A_R\wedge A_R+\\
&+\tau^2\Big[ iA_L\wedge F_R\wedge F_R-iA_R\wedge
F_L\wedge F_L +\frac{i}{2}(A_L-A_R)\wedge
(F_L\wedge F_R+F_R\wedge F_L)+\\
&+\frac{1}{2}A_L\wedge A_L \wedge A_L \wedge F_L-
\frac{1}{2}A_R\wedge A_R \wedge A_R \wedge F_R+
\frac{i}{10} A_L\wedge A_L \wedge A_L\wedge A_L\wedge A_L+\\
&-\frac{i}{10} A_R\wedge A_R \wedge A_R\wedge A_R
\wedge A_R\Big]+\\
&+i\tau^3\,d\tau \wedge\Big[ (A_L\wedge A_R-A_R\wedge A_L)
\wedge (F_L +F_R ) +i A_L \wedge A_L \wedge A_L\wedge A_R+\\
&-\frac{i}{2}A_L\wedge A_R\wedge A_L\wedge A_R +i A_L
\wedge A_R\wedge A_R \wedge A_R \Big]+\\
&\left.+\frac{i}{20}\tau^4 (A_L-A_R)\wedge (A_L-A_R)
\wedge (A_L-A_R)\wedge (A_L-A_R)\wedge (A_L-A_R)\right\}
\end{split}
\ee
We now consider a gauge transformation of (\ref{5-form}).
As we argued above, the tachyon vacuum configuration (\ref{vacuum})
we are considering breaks the flavor invariance to the
diagonal (vectorial) subgroup of $U(N_f)_L\times U(N_f)_R$.
Therefore, taking $V_L=V_R=V$, the gauge
transformation of the fields involved in (\ref{5-form})
reads (see appendix \ref{app: conventions} for more details)
\be\label{ggggg}
\begin{split}
&\delta_\Lambda A_L=-i\, D\Lambda=-i\,d\Lambda -A_L \Lambda+\Lambda A_L\\
&\delta_\Lambda A_R=-i \,D\Lambda=-i\,d\Lambda -A_R\Lambda+\Lambda A_R\\
&\delta_\Lambda F_L =[\Lambda,F_L]\\
&\delta_\Lambda F_R =[\Lambda,F_R]
\end{split}
\ee
Inserting (\ref{ggggg}) in (\ref{5-form}) we find:
\be
\begin{split}
\delta_\Lambda\Omega_5&=\frac{1}{6}\tr\,e^{-\tau^2}\;
d\Lambda\wedge \Big\{(1+\tau^2) \big[-F_L\wedge F_L-
\frac{i}{2}(A_L\wedge A_L\wedge F_L+A_L\wedge F_L\wedge A_L+\\
&+F_L\wedge A_L\wedge A_L)+\frac{1}{2} A_L\wedge A_L\wedge
A_L\wedge A_L+F_R\wedge F_R +\frac{i}{2}(A_R\wedge A_R\wedge F_R+\\
&+A_R\wedge F_R\wedge A_R+F_R\wedge A_R\wedge A_R)-\frac{1}{2}
A_R\wedge A_R\wedge A_R\wedge A_R\big]+\\
&+\tau^3\,d\tau \wedge \big[(A_L-A_R)\wedge (F_L+F_R)+(F_L+F_R)
\wedge (A_L-A_R)+i(A_L\wedge A_L\wedge A_L+\\
&-A_L\wedge A_L\wedge A_R-A_R\wedge A_L\wedge A_L+A_L\wedge A_R
\wedge A_R+A_R\wedge A_R\wedge A_L-A_R\wedge A_R\wedge A_R)\big]\Big\}
\end{split}
\label{dlw5}
\ee
As expected, it is straightforward to show that the
5-form $\delta_\Lambda\Omega_5$ is closed
\be\label{CSclosed}
d \left(\delta_\Lambda\Omega_5\right)=0
\ee
The result (\ref{CSclosed}) ensures that the gauge variation of
the 5-dimensional CS action (\ref{5dCS}) is given by
4-dimensional boundary terms.
The UV boundary term (where $\tau\to 0$)  reproduces the field theory global anomaly \cite{WAdS},
as we will see below.
However, if we consider a background dual to a confining gauge theory,
the gauge variation of the CS action (\ref{5dCS}) may receive a contribution
also from a potential  IR boundary. Because of the confinement property, the space-time
ends smoothly in the IR, and the brane-antibrane world-volume must end before or at the IR end of space.
The contribution to the gauge variation of the CS action
coming from this potential IR boundary
would give rise to an additional contribution to the  gauge anomaly, and make the bulk theory inconsistent. It must therefore vanish.
One is led then to the consistency condition
 $\tau \to \infty$ at the IR, which leaves us  with the UV contribution alone:
\be\label{varCS}
\delta_\Lambda S_{CS} =  (-)^{p} \frac{i N_c}{(2\pi)^5 \a'^3}
 \int_{\mathcal{M}_5}
\delta_\Lambda \Omega_5=\frac{i N_c}{24\pi^2}
\int_{\mathrm{Mink}^4}\tr \,\Lambda(\eta_L -\eta_R)
\ee
where we have used the fact that  the tachyon vanishes on
the UV boundary and reinserted a $(2\pi\a')^3$ factor
in (\ref{dlw5}). We
have defined the 4-form $\eta$ as
\be\label{defeta}
\eta\equiv (-)^{p+1}\left[F\wedge F +\frac{i}{2}(A\wedge A\wedge F+A\wedge
F\wedge A+ F\wedge A\wedge A)-\frac{1}{2}A\wedge A\wedge A\wedge A\right]
\ee
Notice that the computation is rather similar to the  one
in \cite{Sakai:2004cn}, but here we have included the
dependence on the tachyon\footnote{Notice the
different definition of the gauge field with respect to \cite{Sakai:2004cn}.
$A=i\tilde A\,,\ F=i\tilde F$ where $\tilde A,\tilde F$ are the
gauge field and field strength used in \cite{Sakai:2004cn}.}.

In fact, we have checked the anomalous
behavior under a transformation within the $U(N_f)_V$ symmetry
preserved by the vacuum. It is possible to find the anomaly
in the full $U(N_f)_L\times U(N_f)_R$ chiral symmetry before it is
spontaneously broken. For that purpose, we must consider
a more general ansatz in which $T$ is
 proportional to a unitary matrix
$TT^\dagger = T^\dagger T=\tau^2 \II_{N_f}$. With this
milder assumption,
the generalization of (\ref{6form}) reads:
\be\label{6-f}
\begin{split}
\mathrm{Str}&\ e^{i\mathcal{F}}\big|_{6\mathrm{-form}}=
\frac{1}{6}\tr e^{-T^\dagger T}\Big\{- i F_L\wedge F_L
\wedge F_L + i F_R\wedge F_R
\wedge F_R +\\
&+DT^\dagger \wedge DT \wedge F_L \wedge F_L +
DT^\dagger \wedge F_R \wedge DT \wedge F_L -
DT \wedge DT^\dagger \wedge F_R \wedge F_R+\\
&+\frac{i}{4} DT^\dagger \wedge DT \wedge DT^\dagger
\wedge DT \wedge F_L -
\frac{i}{4} DT \wedge DT^\dagger \wedge DT \wedge
DT^\dagger \wedge F_R+\\
&-\frac{1}{60} DT^\dagger \wedge DT
\wedge DT^\dagger \wedge DT \wedge DT^\dagger \wedge DT
\Big\}
\end{split}
\ee
One should now follow the same procedure as above
and generalize the expressions (\ref{5-form})-(\ref{dlw5}).
However, it proves a difficult task to find the five-form
$\Omega_5$. In any case, the IR term will again vanish when
$\tau\to \infty$ due to the overall exponential  factor.
Since in the UV $\tau\to 0$, one can find the UV contribution
even without knowing the full expression for
$\delta_\Lambda \Omega_5$. The result is of course
(\ref{varCS}) but now with independent parameters for the
left and right gauge transformations.
\be\label{varCS2}
\delta_\Lambda S_{CS} = \frac{i N_c}{24\pi^2}
\int_{\mathrm{Mink}^4}\tr
\left[\,\Lambda_L \eta_L -\Lambda_R\eta_R\right]
\ee
Possibly up to a sign, (which is a matter of conventions), (\ref{varCS2})
reproduces the QCD chiral anomaly, as we now check by
using the general procedure sketched in \cite{WAdS} and further
developed in \cite{Freedman:1998tz}. The AdS/CFT conjecture
\cite{Maldacena:1997re,WAdS}
states that $S_{cl}[A]=W[A]$ where $W$ is the generating functional
for current ($J^A=\frac{\delta W}{\delta A^A}$) correlators in the boundary theory. We can equate the
gauge variation of both terms in the equality: $\delta_\Lambda S_{cl}[A]$
is given in (\ref{varCS2}) while
$\delta_\Lambda W[A]=\frac{\delta W}{\delta A^A}\delta_\Lambda A^A=
\int d^4x\, D_\mu\Lambda^A\, J^{A\,\mu}=
-\int d^4x\, \Lambda^A  (D_\mu J^\mu)^A$. The anomalous
divergences of the $U(N_f)_{L,R}$ flavor currents in the presence of sources are,
thus:
\be
\label{anomaly1}
\begin{split}
&\partial_\mu J^{U(1)\,\mu}_{L,R} =(-)^{h_{L,R}}\frac{N_c}{24\pi^2}*\tr\left(\eta_{L,R}\right)\\
&\left(D_\mu J^\mu_{L,R}\right)^a =(-)^{h_{L,R}}\frac{N_c}{24\pi^2}
*\tr\left( \lambda^a \eta_{L,R}\right)
\end{split}
\ee
Here $*$ stands for the four-dimensional Hodge dual,
we have defined $h_L$ and $h_R$ to be $0$ and~$1$
respectively, and the currents have been decomposed in
their abelian and non-abelian components following the conventions of appendix \ref{app: conventions}.
Expressions (\ref{anomaly1}) reproduce the known QCD results (see for example \cite{Jackiw}).

For completeness, we report here the explicit expressions of the traces appearing in equations (\ref{anomaly1}). They can be easily derived from the definition (\ref{defeta}) of $\eta$:
\be
\tr(\eta)= (-)^{p+1}\left(\frac{1}{2}F^a\wedge F^a+N_f \, F^{U(1)}\wedge
F^{U(1)} -\frac{1}{8}f^{ab}_{\phantom{ab}c}\, A^a\wedge A^b\wedge F^c\right)
\ee
and
\be
\begin{split}
\tr(\lambda^a\eta)&= (-)^{p+1}\left[ \frac{1}{2} d^{abc} F^b\wedge F^c +F^a\wedge F^{U(1)}-\frac{1}{8} f^{bc}_{\phantom{bc}a} \left( 3 A^b\wedge A^c \wedge F^{U(1)}-2 A^{U(1)}\wedge A^b\wedge F^c\right)+\right.\\
&\left.-\frac{1}{8}\left(f^{ab}_{\phantom{ab}e} d^{ecd} -f^{ac}_{\phantom{ac}e}d^{ebd}+f^{bc}_{\phantom{bc}e}d^{ead}\right) A^b\wedge A^c \wedge F^d+\frac{1}{16}f^{bc}_{\phantom{bc}h}f^{de}_{\phantom{de}l}d^{ahl} A^b\wedge A^c\wedge A^d\wedge A^e \right]
\end{split}
\ee

\subsection{The $U(1)_A$ axial symmetry}
\label{sect: axial}

The first equation in (\ref{anomaly1}) for the anomalous divergence of
the abelian flavor currents in the presence of external sources  is
not complete. In the case of the $U(1)_A$ symmetry, an
additional anomaly appears  \cite{Adler:1969gk}.

To study this $U(1)_A$ anomaly, we need to know what bulk field couples
to $G\wedge G$ on the boundary (here $G$ denotes the field strength of
the non-abelian color gauge field). We consider then a probe D$q$
color brane. Since we are studying four-dimensional gauge theories, the
D$q$ brane wraps a $(q-3)$-cycle $\mathcal{K}_{q-3}$ of the internal
geometry. Typically, for a background dual to a confining color gauge
theory with no moduli space nor low-energy massive scalars, the D$q$-brane
embedding is stable only at the boundary, in the far UV region.

The action for this probe brane reads:
\be
S_{probe}=-T_q\int_{\Sigma_{q+1}} d^{q+1}y\;e^{-\phi}\sqrt{-\det\left(P[g]+G\right)}+T_q\int_{\Sigma_{q+1}} C\wedge \tr\: e^{i G}
\ee
where $P[g]$ is the pull-back of the background metric on the world-volume $\Sigma_{q+1}$ of the D$q$-brane, and as before we have set $2\pi\a'=1$. The coupling to $G\wedge G$ we are interested in comes from the expansion of the exponential in the  WZ term, and is given by $\frac{T_q}{2}\int_{\Sigma_{q+1}} C_{q-3} \wedge \tr (G\wedge G)$.

Comparison of this term  to the usual Yang-Mills action gives the holographic definition for the QCD $\theta$-angle (recall that $p+q=12$)
\be\label{tym}
\theta_{QCD}=4\pi^2 T_{12-p}\int_{\mathcal{K}_{9-p}} C_{9-p}
\ee
where the integral over $\mathcal{K}_{9-p}$ is evaluated in the far UV (the boundary).

To study effects on the dynamics of flavors related to $\tr\, G\wedge G$, we will need, therefore,  to turn on the RR potential $C_{9-p}$ in the background,
 and look at its interaction with the flavor branes. Using Hodge duality, this potential has the
 same degrees of freedom as $C_{p-1}$. The coupling of this potential to the D$p$-$\overline{\mathrm{D}p}$ flavor branes can then be read
 from  the WZ action (\ref{WZtach})
\be\label{WZ for U(1)A}
S_{WZ}=-i\,T_p \int_{\Sigma_{p+1}}  C_{p-1}\wedge \left. \mathrm{Str} \;e^{i\mathcal{F}}\right|_{2\mathrm{-form}}
\ee
If we take $T$ to be proportional to a unitary matrix,  $T T^\dagger = \tau^2 \II_{N_f}$,  as we did at the end of  subsection \ref{sect: anomaly}, it is easy to write from (\ref{AFdef}) an explicit formula for  $ \left. \mathrm{Str} \;e^{i\mathcal{F}}\right|_{2\mathrm{-form}}$:
\be\label{Str2}
 \left. \mathrm{Str} \;e^{i\mathcal{F}}\right|_{2\mathrm{-form}}= e^{-\tau^2}
\tr (i F_L -i F_R -
DT^\dagger \wedge DT)
\ee

The coupling in  (\ref{WZ for U(1)A}) modifies the equation of motion of $C_{p-1}$, or equivalently the Bianchi identity for the Hodge dual $C_{9-p}$. Indeed, the terms of the overall action involving $C_{p-1}$ are
$S=-\frac{1}{2\kappa_{(10)}^2}
\int \frac12 F_p\wedge *F_p - iT_p \int_{\Sigma_{p+1}}
C_{p-1}\wedge \left. \mathrm{Str} \;
e^{i\mathcal{F}}\right|_{2\mathrm{-form}}$,
from which we obtain
\be\label{dFbar}
d\tilde{F}_{10-p}=d*F_p=2i \kappa_{(10)}^2  T_p\; \delta_{9-p}\left(\Sigma_{p+1} \right)\wedge \left. \mathrm{Str} \;e^{i\mathcal{F}}\right|_{2\mathrm{-form}}
\ee
where $\tilde{F}_{10-p}$ is the gauge-invariant (with respect to world-volume gauge transformations)
$(10-p)$-form RR field strength, and $\delta_{9-p}\left(\Sigma_{p+1} \right)$ is the volume
form for the space transverse to the flavor branes times a $\delta$-function  localized at the position of the flavor branes.

{}From (\ref{dFbar}) we can write:\footnote{The following formula has been computed for the case where all the quarks have the same (possibly vanishing) mass, which is the situation considered in this paper. Since in section \ref{sect: eta'} we will relate $\Omega_1$ to the would-be Goldstone boson $\eta'$, we should expect  that in a more general case $\tr(\log T-\log T^\dagger)$ will be substituted by an expression of the form  $\log(\det T (T^\dagger)^{-1})$. We will not pursue this computation here. }
\be
\tilde F_{10-p} = dC_{9-p} + 2 i (-)^{p+1} \kappa_{(10)}^2  T_p \;\delta_{9-p}\left(\Sigma_{p+1} \right)\wedge  \Omega_1
\label{Fbar}
\ee
where ${\Omega}_1$ is defined by $d{\Omega}_1= \left. \mathrm{Str} \;e^{i\mathcal{F}}\right|_{2\mathrm{-form}}$ and can be explicitly obtained from (\ref{Str2}):
\be\label{Omega1}
\Omega_1 = e^{-\tau^2} \tr \left( i A_L -iA_R
+ (\log T - \log T^\dagger)\tau d\tau \right)
\ee
As we show in appendix \ref{app: gauge transf}, the only gauge transformations under which  $\Omega_1$
transforms non-trivially are the $U(1)_A$ axial ones. We thus consider  an infinitesimal
such transformation $\Lambda_L = -\Lambda_R =i  \alpha \II_{N_f}$:

\be
\delta \Omega_1 =  2 i N_f e^{-\tau^2} (d\a -
\a\ d(\tau^2)) = d\left( 2i  N_f e^{-\tau^2} \a\right)=d\omega_0
\ee
where we defined   $\omega_0 = 2i N_f e^{-\tau^2} \a$

Since $\tilde F_{10-p}$ is gauge invariant,
and as we just showed $\delta_\Lambda {\Omega}_1= d\omega_{0}$,
it follows from (\ref{Fbar}) that the RR-potential cannot be invariant under flavor brane
world-volume gauge transformations \cite{Green:1996dd}:
\be
\delta_\Lambda C_{9-p} =-2i  \kappa_{(10)}^2 T_p
\;\delta_{9-p}\left(\Sigma_{p+1} \right)\; \omega_0\,\,,
\ee
which, inserted in (\ref{tym}) and using the identity $8\pi^2 T_{12-p}\, T_p\, \kappa_{10}^2 =1$ (in units
of $2\pi\a'=1$), results in the following  transformation relation for the QCD theta angle under $U(1)_A$ transformations:
\be
\delta_\Lambda \theta_{QCD} = - i \left. \omega_0\right|_{UV}= 2  N_f \alpha\,\,.
\ee
Because of the boundary coupling $\int d^4x \frac{\theta_{QCD}}{8\pi^2} \tr(G\wedge G )$,  the formula (\ref{anomaly1}) for the divergence of the $U(1)_A$ current $J^{U(1)\,\mu}_A\equiv J^{U(1)\,\mu}_L-J^{U(1)\,\mu}_R$  has to be corrected in order to take into account the non-trivial transformation of $\theta_{QCD}$ (the $U(1)_V$ current $J^{U(1)\,\mu}_V$ is unaffected):
\be\label{anomaly GG}
\partial_\mu J^{U(1)\, \mu}_A=\frac{N_c}{24 \pi^2}*\tr
\left( \eta_L+\eta_R  \right)+ \frac{N_f}{16\pi^2} \;
\epsilon^{\mu_1\ldots\mu_4}\tr_{SU(N_c)}(G_{\mu_1\mu_2} G_{\mu_3\mu_4} )
\ee

\subsection{A holographic view of the Coleman-Witten theorem}
\label{sect: CW}

In \cite{CW}, Coleman and Witten proved that, under a few assumptions, in
$N_c\to \infty$ massless QCD, chiral symmetry is spontaneously broken
$U(N_f)_L \times U(N_f)_R \to U(N_f)_V$.

One of the main results of the previous section is that,
if the theory is confining, the tachyon has to get
a (diverging) vev near the IR even if $T\to 0$ in the UV. Since $T$ transforms in the bifundamental representation of the flavor group,
$\langle T\rangle  \neq 0$ means that the chiral symmetry is broken. We showed
in section \ref{sect: vev} that
 $\langle T\rangle $ is proportional to the identity matrix,
 and consequently the flavor symmetry is broken down
to $U(N_f)_V$.
Confinement, therefore, implies spontaneous chiral symmetry
breaking, reproducing in a holographic language the result of \cite{CW}.

To make the analogy between these two different approaches more evident,
 we can compare the assumptions of \cite{CW} with those we made in the present holographic setup.
In order to do this, we start by  reviewing the five
assumptions of \cite{CW}:
1)~The large $N_c$ limit of QCD exists. 2)~In that
limit the theory is confining.
3)~There is a single order parameter for the breaking
that is a quark bilinear and transforms in the
bifundamental of $U(N_f)_L \times U(N_f)_R$. 4)~Its vev can be found by minimizing
some potential. 5)~The potential is assumed not to have degenerate minima.

For the holographic setup, assumptions 1) and 2)
are also necessary, and of course,
one must further assume the (non-trivial) fact of the
existence of a holographic string
dual of large $N_c$ QCD which fits in the general
framework we have introduced. It is appealing that assumption
3) is automatic in the setup of our work since it is
a well established fact that the lowest state of a
brane-antibrane system is the open
string tachyon. Concerning assumption 4), the string theory
dual provides, in principle, a way to determine the
vev of the quark bilinear:
the bulk vev of the
tachyon field is determined
by solving the bulk field equations with the
UV boundary condition corresponding
to $m_q = 0$ (\ref{UVtau}) and the IR consistency condition
$\tau|_{IR} \to \infty$. Then, one can read the value of
the quark bilinear vev
$\sigma$ from the UV behavior of $\tau(z)$ (\ref{UVtau}).
This
computation requires the DBI
part of the action, and is similar to that
in \cite{Babington:2003vm,Kruczenski:2003uq}. It was
 discussed in section \ref{sect: vev}.

In principle, one expects that the UV and IR
conditions determine uniquely the bulk
vev of the tachyon and therefore the
$\langle \bar q q\rangle $ condensate of the dual theory.
Nevertheless, in the very general
setup we are considering, we cannot prove that the
solution for the $\langle \bar q q\rangle $ is indeed unique, so this
must be taken as a further assumption,
analogous to the non-degeneracy  5) listed above.

We remind the reader that in section \ref{sect: vev}
we also made the (reasonable) assumption that the stable
minimum given by the vacuum (\ref{vacuum}) is the real
vacuum of the theory.

The demonstration presented in this paper, namely  that confinement is
a sufficient condition for spontaneous chiral symmetry breaking,
can be thought of as a reformulation of the geometrical picture of
\cite{Aharony:2006da}. There  it was argued that if one places branes and
antibranes in a background  that smoothly
ends at some point of the radial coordinate
(and therefore is confining), they must necessarily
recombine in the IR, sparking the breaking of chiral symmetry.
 The advantage of the present formulation stands in the fact that the tachyon,
the scalar responsible for the symmetry breaking, is explicitly
taken into account which, for instance, allows to
introduce a bare quark mass.
Notice that, if one is in the conformal window of
${\cal N}=1$ or ${\cal N}=0$ QCD, then there is no IR boundary and
it is consistent to
take a vanishing vev for the tachyon as in
\cite{Klebanov:2004ya,BCCKP}.
However, as in \cite{CW} or \cite{Aharony:2006da}, our argument does not
rule out having spontaneous chiral symmetry breaking without
confinement.

\subsection{The effects of a non-trivial quark mass}
\label{sect: 3point}

In \cite{Freedman:1998tz}, it was explained
how an expression like (\ref{anomaly1}) can be related
to the anomalous three-point function
in the case of ${\cal N}=4$ SYM where the global symmetry
associated to the current $J$ is  $SU(4)_R$.
Analogously, in our case one can relate (\ref{anomaly1}) to the anomalous
massless QCD three-point function.
This anomaly equation in the large $N_c$ limit
was used by Coleman and Witten \cite{CW} to prove the
existence of massless Goldstone bosons, and therefore the
spontaneous breakdown of the symmetry
$U(N_f)_L \times U(N_f)_R \to U(N_f)_V$.
Clearly, if there is a non-zero bare mass for the quarks,
 there
cannot be spontaneous breaking nor massless Goldstone bosons,
so (\ref{anomaly1}) must be modified when $m_q \neq 0$.
In section~\ref{sect: Goldstone}, we will show by analyzing
the equations that define the mesonic spectrum that there are massless
Goldstones if and only if $m_q =0$.
The goal of this section is
to describe holographically the modification of (\ref{anomaly1}).

The axial current is classically non-conserved
when $m_q \neq 0$. In the holographic picture this shows up
because a gauge transformation of the bulk gauge fields is
accompanied by a gauge transformation of the tachyon
(\ref{gaugevar}), which contributes to (\ref{anomaly1}) iff
$m_q \neq 0$. In the rest of
this section we make these statements more
precise, assuming that the space is asymptotically AdS:
\be
g_{xx}(z)\simeq R_{AdS}^2/z^2+\ldots\qquad \quad g_{zz}(z)\simeq R_{AdS}^2/z^2+\ldots
\ee

The antihermitian part of the tachyon matrix $\frac{T-T^\dagger}{2}$
is dual to ($-i$ times) the pseudoscalar current,
see table 1. We consider a perturbation around the
vacuum $T=\tau e^{i \theta}$ where $\theta$ is proportional to the (hermitian)
pion matrix, $\theta\propto\frac{\eta'}{\sqrt{2N_f}}\II+  \pi^a\lambda^a$.
At first order, \mbox{$\frac{T-T^\dagger}{2}= i \tau \theta $}.
 There must exist a boundary
coupling contributing in the $W$ generating functional
$\int d^4x\,(- i \phi_0^A   J_5^A)_{z=0}$, where $J_5$ is the pseudoscalar (pion)
current $J_5 =i\bar q \gamma^5 q+
 i\bar q \gamma^5 \lambda^a q$ and
$\phi_0$ is proportional to
$i \tau\theta$.
As argued in section~\ref{sect: vev}, we require that the mass
of the tachyon, at
least near the boundary, is $m_T^2 R_{AdS}^2 = -3$
since $\bar q q$ and $i\bar q \gamma^5 q$ are operators of dimension 3.
Following Witten's prescription
\cite{WAdS}, one has to couple to the pseudoscalar current
in the boundary,
 $\phi_0 = R_{AdS}^\frac32 \frac{1}{z} i\tau_{can}  \theta$,
since for a scalar of the above cited mass, the non-normalizable
behavior is $\sim z$. Inserting the value of
$\tau_{can}$ given in~(\ref{UVtau}), the contribution to the generating
functional is:
\begin{equation}
\int d^4 x \, (m_q\,\theta^A J_5^A)
\label{thcoupling}
\end{equation}
We now compute how this expression transforms under an
axial gauge transformation $\Lambda_L = -\Lambda_R = \Lambda=i\Lambda^A\lambda^A=i\alpha \II+i\Lambda^a \lambda^a$.
{}From (\ref{gaugevar}), the infinitesimal gauge transformation
$V\equiv e^{\epsilon \Lambda}$ of the tachyon around the vacuum
(\ref{vacuum}) is given by
$\delta_\Lambda T = \Lambda_R T - T \Lambda_L =
\tau (\Lambda_R-\Lambda_L) =- 2\Lambda \tau$, which in turn can
be expressed as a transformation of the pion matrix $\theta$ as:
$\delta_\Lambda \theta^A = (2 i \Lambda)^A=-2\Lambda^A$, which, inserted in
(\ref{thcoupling}), yields $\int d^4 x \, m_q\, (-2\alpha J^{U(1)}_5-2\Lambda^aJ_5^a)$.
This leads to a modification of (\ref{anomaly1}) and (\ref{anomaly GG}):
\be
\begin{split}
&\partial_\mu J^{U(1)\, \mu}_A=-2m_q J^{U(1)}_5 + \frac{N_c}{24\pi^2} *\tr\left( \eta_L+\eta_R  \right)
+ \frac{N_f}{16\pi^2} \;
\epsilon^{\mu_1\ldots\mu_4}\tr_{SU(N_c)}(G_{\mu_1\mu_2} G_{\mu_3\mu_4} ) \\
&\left(D_\mu J^\mu_A\right)^a =-2m_q J_5^a +\frac{N_c}{24\pi^2} *\tr\left(\lambda^a (\eta_L+\eta_R)\right)
\end{split}
\label{anomalouswithmass}
\ee
One can now relate these expressions to the three-point function
in the presence of a non-trivial quark mass. The term with
$m_q$ in (\ref{anomalouswithmass})
results in an extra term for the  three-point function.
The presence of this new term
invalidates the  argument that Coleman and Witten
used in the $m_q=0$ case to show that the three-point function
has a pole at zero momentum.
Thus, as expected, the three-point function can be analytic
at zero momentum,  consistent with the absence of
massless Goldstone bosons.

\section{Features of the mesonic mass spectrum}
\label{sect: DBI}

\setcounter{equation}{0}

\subsection{Equations determining the spectrum}
\label{sect: excitations}

We now study small fluctuations of the different
fields around the vacuum configuration.
They correspond to
the mesons of the field theory.
We will just consider quadratic terms in the action, which
are enough to find the mass spectrum, but not the couplings.
At this level, the non-abelian action (\ref{generalact2}) is just the
sum of $N_f^2$ copies of the abelian one~(\ref{generalact}).
Thus, we use (\ref{generalact}) in the following
and there are $N_f^2$ copies of each of the towers of resonances that we
will study.

We start by
considering excitations for the fields $\theta, A^L, A^R$
around the vacuum of section \ref{sect: vev}. Fluctuations of the modulus of
the tachyon $\tau$ will be briefly mentioned below. We expand the
simplified version of (\ref{generalact}) (neglecting the
scalars $Y_{(i)}^I$ and assuming there is no $B$-field).
We define:
\be
e^{-\tilde\phi}=e^{-\phi_{eff}} V(\tau^2)\,,\qquad
\tilde g_{zz} = g_{zz} + \frac{2}{\pi} (\partial_z \tau)^2
\label{somedefs}
\ee
In order to use a notation similar to the one in
\cite{Erlich:2005qh}, we choose a gauge:
\be
A_z^L= A_z^R=0
\ee
and define:
\bear
V_M = \frac{A_M^L+A_M^R}{2}\,,\qquad
A_M = \frac{A_M^L-A_M^R}{2}\,,\qquad
v = \frac{2\sqrt2}{\sqrt{\pi}}\tau\,,
\label{defs2}
\eear
and $V_{\m\n},A_{\m\n}$ as the (abelian) field strengths of
$V_{\m},A_{\m}$.
Expanding (\ref{generalact}), we find:
\bear
S&=&-\int d^4x dz e^{-\tilde \phi} \left[
\frac12 \tilde g_{zz}^\frac12 (V_{\m\n}V^{\m\n}+
A_{\m\n}A^{\m\n})+ g_{xx} \tilde g_{zz}^{-\frac12}
\left((\partial_z V_\m)^2+(\partial_z A_\m)^2\right)+\right. \rc
&&+\left.
\frac14 g_{xx}  \tilde g_{zz}^\frac12  v^2 (\partial_\m \t
+ 2 A_\m)^2 +
\frac14
 g_{xx}^2  \tilde g_{zz}^{-\frac12} v^2 (\partial_z \t)^2\right]
\label{lagr2}
\eear

\paragraph{Vector sector}

\

\noindent
The vector ($1^{--}$) sector is decoupled from the rest
in the action (\ref{lagr2}). The equation of motion for $V_\m$
can be solved by  expanding $V_\m$ in modes:
\be
V_\m = \sum_n \psi_{(n)}(z) {\cal V}_\m^{(n)} (x^\m)
\label{vectorexpand}
\ee
with:
\be
\partial_z ( e^{-\tilde\phi} g_{xx} \tilde g_{zz}^{-\frac12}
\partial_z \psi_{(n)} ) + m_n^2 e^{-\tilde \phi} \tilde g_{zz}^{\frac12}
\psi_{(n)} = 0
\label{eqvector}
\ee
The physical modes are those which yield a finite action
when integrating in $z$.
Substituting (\ref{vectorexpand}) in (\ref{lagr2}),
one finds a tower of massive vectors:
\be
S = -\int d^4x \sum_n \left[ \frac12 {\cal V}_{\m\n}^{(n)} {\cal V}^{\m\n}_{(n)}
+ m_n^2  {\cal V}_{\m}^{(n)} {{\cal V}^{\m}}^{(n)}\right]
\ee
provided the normalization
conditions:
\footnote{\label{foot1}
We have chosen this normalization such that when the
fields are promoted to their non-abelian generalization
${\cal V}_{\m}^{(n)} \to{\cal V}_{\m}^{(n)\;a}\lambda^a$,
the components ${\cal V}_{\m}^{(n)\;a}$ are normalized in
the standard way, since we are using
$\tr \lambda^a \lambda^b=\frac12 \delta^{ab}$.
We apply this prescription to all the different towers of modes.}
\bear
\int_0^{z_{IR}} dz \frac12 e^{-\tilde\phi}
\tilde g_{zz}^\frac12 \psi_{(n)}^2 &=& \frac12
\rc
\int_0^{z_{IR}} dz e^{-\tilde\phi} g_{xx} \tilde g_{zz}^{-\frac12}
(\partial_z \psi_{(n)})^2 &=&  m_n^2
\label{norm1}
\eear
Integrating by parts the second expression and
using (\ref{eqvector}),
we consistently obtain the first line of (\ref{norm1}).

\paragraph{Axial vector sector}

\

\noindent
The vector field fluctuation $A_\m$ can be split in a
transverse and a longitudinal part,
$A_\m = A^\bot_\m + A^\Vert_\m$, with
$\partial^\m A^\bot_\m =0$. We first consider the transverse part,
corresponding to $1^{++}$ excitations.
We expand it   as:
\be
A^\bot_\m = \sum_n A_{(n)}^\bot (z) {\cal B}_\m^{(n)} (x^\m)
\label{axiexpand}
\ee
which results in a tower of massive axial vectors:
\be
S = -\int d^4x \sum_n \left[
\frac12 {\cal B}_{\m\n}^{(n)} {\cal B}^{\m\n}_{(n)}
+ m_n^2  {\cal B}_{\m}^{(n)} {{\cal B}^{\m}}^{(n)}\right]
\ee
subject to the normalization conditions:
\bear
\int_0^{z_{IR}} dz \frac12 e^{-\tilde\phi} \tilde g_{zz}^\frac12
{A_{(n)}^\bot}^2 &=& \frac12
\rc
\int_0^{z_{IR}} dz \left[e^{-\tilde\phi} g_{xx} \tilde g_{zz}^{-\frac12}
(\partial_z A_{(n)}^\bot)^2
+e^{-\tilde\phi} g_{xx} \tilde g_{zz}^{\frac12}
v^2 {A_{(n)}^\bot}^2\right]
&=& m_n^2
\label{norm2}
\eear
and the second order differential equation:
\be
\partial_z (e^{-\tilde\phi} g_{xx} \tilde g_{zz}^{-\frac12} \partial_z
A^\bot_{(n)})
+ m_n^2 e^{-\tilde\phi} \tilde g_{zz}^{\frac12}A^\bot_{(n)}-
 e^{-\tilde\phi} g_{xx} \tilde g_{zz}^{\frac12} v^2 A^\bot_{(n)} =0
 \label{eqaxial}
\ee
It is interesting to notice at this point a difference with
respect to constructions like the Sakai-Sugimoto model
\cite{Sakai:2004cn}. In that kind of models, the vector
and axial vector mesons are solutions of a single second order
differential equation with different matching conditions at the IR.
In this case, they satisfy different equations
(\ref{eqvector}), (\ref{eqaxial}). If the chiral symmetry were not
broken ($v=0$), the equations would become degenerate and vector and axial
vector mesons would have the same mass spectrum due to the
unbroken symmetry.
The differential equations found in this construction coincide with
those of the AdS/QCD models  \cite{Erlich:2005qh,DaRold:2005zs}
up to the $z$-dependent dilaton, metric factors and tachyon vev,
which we have kept general.

\paragraph{Pseudoscalar sector}

\

\noindent
The modes coming from $\t$ and the longitudinal part of $A_\m$ combine
to give a single tower of resonances with the
quantum numbers of pions $0^{-+}$.
The equations of motion from (\ref{lagr2}) can be solved by expanding:
\bear
A^\Vert_\m &=& -\sum_n \varphi_{(n)}(z) \partial_\m(\alpha^{(n)} (x^\n))\rc
\theta &=& 2\sum_n \vartheta_{(n)}(z)\alpha^{(n)} (x^\n)
\label{pionexpand}
\eear
where the functions $\varphi_{(n)}$ and $\vartheta_{(n)}$ satisfy the following coupled differential equations
\bear
\partial_z (e^{-\tilde\phi} g_{xx}
\tilde g_{zz}^{-\frac12} \partial_z \varphi_{(n)})
+ e^{-\tilde\phi} g_{xx} \tilde g_{zz}^{\frac12} v^2
(\vartheta_{(n)} - \varphi_{(n)})
=0 \rc
g_{xx} v^2 \partial_z \vartheta_{(n)} - m_n^2 \partial_z \varphi_{(n)}
=0
\label{eqpion}
\eear
Inserting (\ref{pionexpand}) in (\ref{lagr2}) and normalizing:
\bear
\int_0^{z_{IR}} dz \left[ e^{-\tilde\phi} g_{xx} \tilde g_{zz}^{-\frac12}
(\partial_z \varphi_{(n)})^2+
 e^{-\tilde\phi} g_{xx} \tilde g_{zz}^{\frac12}
v^2 (\vartheta_{(n)} - \varphi_{(n)})^2\right] &=& 1
\label{norm3a}
\\
\int_0^{z_{IR}} dz \left[e^{-\tilde\phi} g_{xx}^2 \tilde g_{zz}^{-\frac12} v^2
(\partial_z \vartheta_{(n)})^2
\right]
&=& m_n^2
\label{norm3}
\eear
one finds
a tower of pseudoscalars in the four-dimensional theory:
\be
S = -\int d^4x \sum_n \left[  (\partial_\m \alpha^{(n)})^2
+ m_n^2  (\alpha^{(n)})^2\right]
\label{pseudoaction}
\ee

\paragraph{Scalar sector}

\

The scalar mesons ($0^{++}$)
 come from excitations around the vacuum of the form
$\delta \tau =  S(x_\m,z)$. Expanding
(\ref{tachyonac})
up to quadratic order in $S$, one finds an action:
\bear
S&=&-2\int d^4x dz e^{-\tilde \phi}
\Big( \frac12 \frac{\partial_\tau^2 V}{V}\Big|_{\tau =v/2}
g_{xx}^2 \tilde g_{zz}^\frac12 S^2+
\frac{\sqrt2}{2\sqrt{\pi}} \frac{\partial_\tau V}{V}\Big|_{\tau =v/2}
g_{xx}^2 \tilde g_{zz}^{-\frac12}(\partial_z v) S (\partial_z S)
+  \rc
&+&\frac{1}{\pi} g_{xx}^2 \tilde g_{zz}^{-\frac32}
(\tilde g_{zz}-\frac14(\partial_z v)^2)
(\partial_z S)^2 + \frac12 g_{xx}
\tilde g_{zz}^\frac12(\partial_\m S)^2
\Big)
\label{eqscalar}
\eear
from which one can straightforwardly extract the linear equation of motion
and normalizability condition. Since we will not need them in the following,
we do not present them explicitly.

\subsection{On ``linear confinement" (for highly excited mesons)}
\label{sect: linear}

In a theory with linear confinement such as QCD, one expects that
the squared masses $m_n^2$ of highly excited
hadrons grow linearly with the excitation number $n$: $m_n^2 \propto n$.
This behavior is difficult to find from holographic models, that
typically yield $m_n^2 \propto n^2$ \cite{Schreiber:2004ie}.
In \cite{Karch:2006pv} it was shown that, by appropriately tuning
the behavior of the metric and dilaton in the IR, one can indeed
find the expected behavior.
Moreover, it was conjectured that such behavior could come from
{\it closed} string tachyon condensation,
but no concrete model producing such
IR asymptotics of  metric and dilaton has been found.

We will show here that the relation $m_n^2 \propto n$ is automatic in our
construction. Therefore, the linear confinement relation for mesons comes from
{\it open} string tachyon condensation on
the flavor brane world-volume.
In this section we just state
the ideas and results while details are relegated to appendix
\ref{sect:WKB}.

The argument relies on the fact that the meson excitations feel the  effective
{\it open string}
dilaton and metric (see eq. (\ref{somedefs}))
rather than just the closed string dilaton and metric.
In the IR, $\tau \to \infty$ so
$\tilde g_{zz}=g_{zz}+\frac{2}{\pi} (\partial_z \tau)^2
 \approx {\frac{2}{\pi} (\partial_z \tau)^2}$
even if $g_{zz}$ has a single pole. Moreover, $g_{xx}$ goes to
a constant, which, on general grounds,
can be identified with the QCD string tension.
Reinserting $\a'$ and  defining a new
radial variable $u=\sqrt{\frac{2}{\pi T_{QCD}}} \tau(z)$,
we have, near the IR, where $u,\tau \to \infty$:
\bear
&&ds^2 = 2\pi\alpha' \left(T_{QCD} dx_{1,3}^2 +
\frac{2}{\pi}(\partial_z \tau)^2 dz^2 \right)=
2\pi\alpha'  T_{QCD} \left(
 dx_{1,3}^2 + du^2 \right)\rc
&&e^{-\tilde \phi} \sim V(\tau^2)\sim
e^{-\tau^2} \sim e^{-\frac{\pi T_{QCD}}{2}u^2}
 \,\, \Rightarrow
\tilde \phi \sim \frac{\pi T_{QCD}}{2}u^2
\label{linbeh}
\eear
It turns out that  the natural radial
variable in the IR is proportional to $\tau$.
The quadratic growth of the dilaton
(\ref{linbeh}) is the IR behavior advocated in \cite{Karch:2006pv}
to account for linear confinement.
In fact, using the
WKB approximation to compute the masses of highly excited bound states
one obtains, for the
vector meson tower (see appendix \ref{sect:WKB}):
\be
m_n^2 \approx 2\pi T_{QCD}\ n \qquad \textrm {(large $n$,
vector mesons)}
\label{linconfrel}
\ee

This is a quite general result that only relies on having a confining
background and therefore diverging tachyon, on the
large $\tau$ gaussian behavior of the tachyon potential and on the
DBI action. We find very encouraging that our
construction naturally accounts for the physically expected relation
of linear confinement, including the correct multiplicative factor.

However, there is an important caveat: if one repeats the same computation for
axial vector mesons, eq. (\ref{eqaxial}), one finds:
\be
m_n^2 = 2\pi \sqrt{1+\frac{16}{\pi^2}}  T_{QCD}\ n
\qquad \textrm {(large $n$,
axial vectors)}
\label{wrong}
\ee
This cannot be physically correct since vector and axial mesons should
asymptotically have equal masses at large $n$, as chiral symmetry is restored
for excited hadrons \cite{Shifman:2005zn}. The problem may arise from the fact that we are using
the DBI action outside its range of validity. This point definitely
 deserves a better understanding.

Finally, it is important to stress that
in view of the confining IR behavior,
there is no need to impose an
arbitrary IR boundary condition. The IR condition
for physical excitations is simply the normalizability of the action
\cite{Karch:2006pv}.

\subsection{Goldstone bosons}
\label{sect: Goldstone}

Large $N_c$
QCD with massless quarks has a set of $N_f^2$ massless pseudoscalars
which are
the Goldstone bosons of the spontaneous breaking of the $U(N_f)_A$.
We will generically call them pions.
In this section we show how they
appear in this construction\footnote{
One of the $N_f^2$ pseudoscalars is massless only in the strict
$N_c \to \infty$ limit. This is described in
 section~\ref{sect: eta'}.}.
They are the solutions to equations (\ref{eqpion})-(\ref{norm3})
with $m_n = 0$. Notice that, as expected, there cannot be massless
(axial) vectors due to (\ref{norm1}), (\ref{norm2}).

We generalize the analysis in
(\ref{pionexpand})-(\ref{pseudoaction}) with $m_n=0$
to the non-abelian $N_f>1$ case.
Define the pion matrix  as the generalization of the
$\alpha^{(0)}(x^\nu)$ in (\ref{pionexpand}):
\be
\pi(x^\nu) = \frac{\eta'(x^\nu)}{\sqrt{2N_f}} \II + \pi^a(x^\nu) \lambda^a
\label{pidef}
\ee
To lowest order, the action for the pions is just the one for a set
of massless scalars:
\be
S=-\int d^4x \tr \left (\partial_\mu \pi \partial^\mu \pi \right)=
-\int d^4x \left (\frac12 \partial_\mu \eta' \partial^\mu \eta'
+ \frac12 \partial_\mu \pi^a \partial^\mu \pi^a \right)
\ee
Due to (\ref{eqpion}), $\vartheta_{(0)}$ must be a constant.
We will shortly see that it is related to the pion decay constant.
We define a quantity:
\be
\xi(z) = \varphi_{(0)}(z) - \vartheta_{(0)}
\ee
which, regarding (\ref{eqpion}), (\ref{norm3a}), satisfies:
\bear
\partial_z (e^{-\tilde\phi} g_{xx} \tilde g_{zz}^{-\frac12} \partial_z
\xi(z))
- e^{-\tilde\phi} g_{xx} \tilde g_{zz}^{\frac12} v^2 \xi(z) =0\,,
\label{xieqa}
\\
-e^{-\tilde\phi} g_{xx} \tilde g_{zz}^{-\frac12}
\xi(z) \partial_z \xi(z) \big|_{z=0} = 1
\label{xieqb}
\eear
For the second line, we have integrated by parts (\ref{norm3a}) and
taken into account that for $\xi(z)$ we must choose the normalizable mode in
the IR.

The question of whether there are massless modes boils down
to the existence of a function $\xi(z)$ satisfying (\ref{xieqa}),(\ref{xieqb}).
{}From (\ref{xieqb}), it is obvious that
the answer depends on the UV behavior of $\xi(z)$, which,
due to (\ref{xieqa}) depends on the UV behavior of
$v=\frac{2\sqrt2}{\sqrt{\pi}}\tau$.

We now give a heuristic
argument suggesting that if quarks are massless, there
are indeed massless Goldstone bosons. Since $\tau$ is dual
to $\bar q q$, the UV behavior of $\tau$ is related to
the quark mass and condensate. If there is no
quark mass, $\tau$ is normalizable, {\it i.e.} it vanishes
fast enough in the UV. It is natural to think then that the second
term in (\ref{xieqa}) can be neglected near the UV, allowing for
 a solution
to (\ref{xieqa}) and (\ref{xieqb}),
which is, asymptotically,
\be \xi(z) = -\vartheta_{(0)} +
\vartheta_{(0)}^{-1} f(z)
\label{xiasympt}
\ee
The constant $\vartheta_{(0)}$ has to be
determined by the IR condition, we have defined $f(z)$ such that
$\partial_z f(z)=(e^{-\tilde\phi} g_{xx} \tilde
g_{zz}^{-\frac12})^{-1}$ and  used the residual gauge invariance to fix $\varphi_{(0)}(0)=0$ and therefore $f(0)=0$.
If the quarks are, instead, massive, $v$ is larger in the UV and the second
term in (\ref{xieqa}) cannot be neglected, which hinders the
existence of a
solution as the one just described corresponding to a massless Goldstone.

Just as a clarifying example, we consider a five-dimensional
space which is asymptotically AdS. In this case, all  expressions
asymptote in the UV
to those of \cite{Erlich:2005qh}, where  one can explicitly
check that the heuristic reasoning above is valid.

\subsection*{The pion decay constant}

Consider $m_q=0$ so that there are indeed massless pions.
The pion decay constant
can be related to the pole of the axial current two-point function
at zero momentum:
\be
\Pi_A (q^2) = \sum_n \frac{f_{A_n}^2}{q^2+ M_{A_n}^2} + \frac{f_\pi^2}{q^2}
\ee
One can compute this correlator following the AdS/CFT
prescription, as in  \cite{Erlich:2005qh,DaRold:2005zs}.
This is done by computing the on-shell action giving
the appropriate boundary condition to the field
dual to the axial current, {i.e} $A_\mu^\bot$.
When $q^2=0$, the bulk equation for $A_\mu^\bot(z)$ is (\ref{eqaxial}):
\be
\partial_z (e^{-\tilde\phi} g_{xx} \tilde g_{zz}^{-\frac12} \partial_z
A^\bot_{(n)}) -
 e^{-\tilde\phi} g_{xx} \tilde g_{zz}^{\frac12} v^2 A^\bot_{(n)} =0
 \label{aboteq}
\ee
Substituting in (\ref{lagr2}) and deriving twice with
respect to ${\cal B}_\m^{(n)}$ (as defined in (\ref{axiexpand})),
one gets:
\be
f_\pi^2 =\lim_{\epsilon \to 0}
\left( - e^{-\tilde \phi} g_{xx} \tilde g_{zz}^{-\frac12}
A^\bot(z)
\partial_z A^\bot(z)|_{z=\epsilon} \right)
\label{fpi}
\ee
where $A^\bot(z)$ is a solution to eq (\ref{aboteq})
subject to the UV condition $A^\bot|_{z=\epsilon}=1$ and the IR
normalizability condition. Notice there is an
extra factor of 1/2 in
(\ref{fpi}) with respect to what one would get from
(\ref{lagr2}). This is again because we use conventions suitable
for the non-abelian generalizations of the fields, see footnote
\ref{foot1}.

Two remarks are in order:
even if the $f_\pi$
in (\ref{fpi}) depends on quantities at $z=0$, the value of
$f_\pi$ depends on the full metric, since one has to select the
well behaved IR mode. This typically involves a non-trivial numerical
integration. Notice also that $f_\pi^2$ is of order $N_c$ since
inside $e^{-\tilde \phi}$ there is a D-brane tension
which scales as $g_s^{-1} \sim N_c$.

Comparing (\ref{xieqa}), (\ref{xieqb}), (\ref{aboteq}),
(\ref{fpi}) and using the fact that both $\xi(z)$ and $A^\bot(z)$
must follow the normalizable behavior in the IR, it is straightforward
to conclude that up to an unimportant sign,
$\xi(z) = -f_\pi^{-1} A^\bot(z)$. This equality at $z=0$ yields:
\be
\vartheta_{(0)} = \frac{1}{f_\pi}
\label{varthetavalue}
\ee

\subsection*{The Gell-Mann-Oakes-Renner relation}

We now show how, considering asymptotically AdS,
one can obtain the
GOR relation \cite{Gell-Mann:1968rz}, which gives the masses of
the pions when the quark mass is small but non-vanishing.
The argument is very similar to the one in \cite{Erlich:2005qh}
but details are different.
In the following we assume $m_q\neq 0$ but only keep
terms linear in $m_q$.

We have to solve (\ref{eqpion})-(\ref{norm3}) for small
$m_n \equiv m_\pi$. This solution can be obtained as a perturbation
of the $m_n=0$ case studied at the beginning of this section.
As opposed to that case, $\vartheta$ cannot be a constant and,
from (\ref{norm3}), one can show that $\vartheta|_{z=0}=0$ is needed.
Then, using (\ref{eqpion}):
\be
\vartheta(z)= m_\pi^2 \int_0^{z}
dz \frac{\partial_{z}\varphi (z)}{g_{xx} v^2}=
\frac{m_\pi^2 f_\pi}{2} \int_0^z dz \frac{z^3}{(m_q z + \sigma z^3)^2}
\ee
For the second equation we have used that the integral is dominated
by the small $z$ region so we can substitute the asymptotic value
of the different functions. To obtain $\partial_{z}\varphi (z)$
we have used its value in (\ref{xiasympt}) and for $v^2$
we have substituted (\ref{defs2}) and (\ref{UVtau}).
As a consistency check, notice that the region where the
integral above has significant support is around
$z\sim \sqrt{m_q/\sigma}$, so taking $m_q$ small enough, the
integral only probes the asymptotically AdS region.

For $z\gg \sqrt{m_q/\sigma}$, the function $\vartheta(z)$ goes to
a constant which has to be the one of the massless case
(\ref{varthetavalue}). Putting everything
together, we get the known expression:
\be
m_\pi^2 = \frac{4 m_q \sigma}{f_\pi^2} =
- \frac{2 m_q \langle \bar qq \rangle}{f_\pi^2} \,,
\qquad  ( m_q \to 0 )
\ee
where we have substituted (\ref{sigmacond}).

\subsection{The mass of the $\eta'$}
\label{sect: eta'}

We now return to the $m_q=0$ case, where, in principle there
are $N_f^2$ massless Goldstones.
In section \ref{sect: axial} it was shown how our setup
correctly reproduces the
$\mathrm{O}\left(\frac{N_f}{N_c}\right)$ anomaly of
the $U(1)_A$ axial symmetry \cite{Adler:1969gk}. This
 anomaly implies that  the
(generalization of the)
$\eta'$ meson, the would-be Goldstone boson corresponding
to the diagonal $U(1)_A$
subgroup of the spontaneously broken $U(N_f)_A$, has a mass of order
$\frac{N_f}{N_c}$, and is therefore massless only
in the strict $N_c \to \infty$ limit.

The $\eta'$  mass appears  in the
present holographic setup via a Stuckelberg mechanism.
The reasoning we will follow
is very similar to the one of \cite{Sakai:2004cn} (see
\cite{Barbon:2004dq,Armoni}
for related work and \cite{Bergman:2006xn} for a recent
discussion in deconfined theories with broken chiral symmetry).

The $C_{9-p}$  can
only appear in the action in the gauge invariant combination
(\ref{Fbar}). Integrating this expression in the
space composed of the $9-p$ non-Minkowski dimensions that the
color branes wrap plus the radial direction we obtain:
\be\label{etaprime}
(2\pi)^2 T_{12-p} \int_{{\cal M}_{10-p}} \bar F_{10-p}= \theta_{QCD}
 + i (-)^{p+1} \int (\Omega_1)_z dz
= \theta_{QCD} +(-)^p  \frac{\sqrt{2N_f}}{f_\pi}\eta'\,\,.
\ee
We have used (\ref{tym}) and
$\int_{{\cal M}_{10-p}} d C_{9-p}= \int_{\mathcal{K}_{9-p}} C_{9-p}$
where the last integral is evaluated in the UV.
For a 3+1 confining theory defined on D$q$-branes with $q\geq 4$
as in \cite{Witten:1998zw,Maldacena:2000yy} this
equality holds  automatically  because the space closes off smoothly at the IR
and $\mathcal{K}_{9-p}$ is the boundary of ${\cal M}_{10-p}$.
On the other hand, if one wants to build some kind of
five-dimensional model in the spirit of
\cite{Erlich:2005qh,DaRold:2005zs,Shock:2006qy,Karch:2006pv,UmutElias},
one would have  to impose that the RR-potential vanishes in the IR
boundary to find the same condition.

The integral in (\ref{etaprime}) is computed substituting
$T=\tau e^{i \theta} = \tau e^{i 2 \vartheta \pi(x^\n)}=
\tau e^{i \frac{2 \pi(x^\n)}{f_\pi}}$ in (\ref{Omega1}):
\be
\int (\Omega_1)_z dz=\int e^{-\tau^2}\tr\left((\log T-\log T^\dagger)\tau
d\tau\right) =
\frac{4i\,}{f_\pi}  \tr (\pi)
\int_0^\infty e^{-\tau^2} \tau d\tau
=i\frac{\sqrt{2N_f}}{f_\pi}\eta'
\label{Omega1int}
\ee
The contribution of $\theta_{QCD}$ to the vacuum energy density
can be computed by integrating the kinetic term of the
RR $C_{9-p}$-form \cite{Witten:1998uk}, and must appear through
the gauge invariant combination (\ref{etaprime}). Thus:
\be
S = -\frac{\chi}{2}
\int d^4x  \left ( \theta_{QCD} + (-)^p
\frac{\sqrt{2N_f}}{f_\pi}\eta'
\right)^2
\label{topolog}
\ee
where the topological susceptibility of the vacuum is, by
definition, the second derivative of the vacuum energy density
in the glue theory without flavors $\chi = \left(\frac{d^2{\cal E}}
{d\theta_{QCD}^2}
\right)_{{\textrm {no quarks}}}$. It can be
computed from the supergravity action \cite{Witten:1998uk}.
The expression (\ref{topolog}) reproduces the Veneziano-Witten formula
for the mass of the $\eta'$
\cite{Witten:1979vv}
\footnote{The difference in a factor of 2 with respect to
\cite{Witten:1979vv} is due to
 a factor $\sqrt2$ in the definition of $f_\pi$. In
our conventions
$\langle 0 | J^{U(1) \,\mu}| \eta' \rangle = \sqrt{2 N_f} p^\mu f_{\eta'}$
and $f_{\eta'} = f_\pi$ up to suppressed ${\cal O}(N_c^{-1})$
corrections.
}:
\be
m_{\eta'}^2 = \frac{2N_f}{f_\pi^2} \chi
\ee

\section{Comments and discussion}

Compared to the  models
\cite{Babington:2003vm} (\cite{Kruczenski:2003uq})
where flavors where introduced with D7 (D6)-branes in the background
of D3 (D4)-branes, our construction has the
advantage that the breaking of the full non-abelian chiral symmetry
is described. In fact, one may think of the  models presented in those
papers as the result
of a D$p\,$-$\overline{{\rm D}p}$ system in which the tachyon has fully
condensed leaving a D$(p-2)$ brane as a vortex. Since a
non-trivial tachyon
breaks the $U(N_f)_A$, this is not present in the final configuration
and only a remnant $U(1)_A$ is left as the rotation symmetry around
the vortex (for a related discussion, see \cite{Sugimoto:2004mh}).

The model of Sakai and Sugimoto \cite{Sakai:2004cn,SS2}
 and its generalizations
share many properties with our setup, but there are also important
differences. In \cite{Sakai:2004cn,SS2}, it does not seem
possible to include a bare mass for the
quark, so clearly there are aspects of the QCD symmetry
breaking that are not present in that setup.
 Another phenomenon that we have shown
to be intimately related to the physics of the tachyon is the existence
of massive resonances with the quantum numbers of pions $0^{-+}$.
These are not present in  the model  of \cite{Sakai:2004cn,SS2}.
A heuristic way of understanding the differences is that the
description of \cite{Sakai:2004cn,SS2} is done already in the broken
phase (with the brane-antibrane pair reconnected), so that information about
the order parameter is lost.
It would be interesting to try to include the tachyon in such
type of models,
generalizing the formalism of this paper
to non-overlapping brane-antibrane with excited transverse scalars.
For some progress in this direction, see \cite{Antonyan:2006vw}.

In a sense, the model presented in this paper can be thought of
as a string theory construction in which the AdS/QCD models
\cite{Erlich:2005qh,DaRold:2005zs}
are embedded.
We have seen that the five-dimensional
field spectrum is the same. Expanding the square roots
for small values of $A_M$ and  $T$, one gets the same kind of action as in
\cite{Erlich:2005qh,DaRold:2005zs}.
In fact, the equations to determine the
spectrum (\ref{eqvector}), (\ref{eqaxial}), (\ref{eqpion})
reduce to those in \cite{Erlich:2005qh,DaRold:2005zs} if
one considers AdS metric and constant dilaton.
On the other hand, the equation for the scalar mesons (\ref{eqscalar})
is different from its analog in \cite{Pomarol2}.
Small $T$, as argued above,
corresponds to the UV. The successful features
of \cite{Erlich:2005qh,DaRold:2005zs} come from the UV. We therefore conclude
that a model built along the lines described in this paper
in asymptotically AdS space can
capture the good physical features of \cite{Erlich:2005qh,DaRold:2005zs}
while the IR arbitrariness of those models is lifted due to the
condition $\tau|_{IR} \to \infty$. This condition fixes the quark
condensate in terms of the quark mass and removes the extra
non-physical parameter appearing in \cite{Erlich:2005qh,DaRold:2005zs}.
Moreover, the brane physics automatically provides the 5d
Chern-Simons term.

By applying the formalism described in this paper to a concrete
model ({\it i.e.} to particular expressions of the metric and dilaton),
one could find numerical estimates of QCD observables
such as meson masses or couplings.
Nevertheless, as stressed several times,
enforcing a well behaved asymptotically AdS space,
implies $m_T^2 R_{AdS}^2 = -3$
because the quark bilinear has dimension 3. Since $m_T^2 =  -\frac{1}{2\a'}$,
we need $R_{AdS}^2 = 6 \a'$ and the space has large curvature, of the
order of the string scale.
The complete dual background, then, is not a solution to just Einstein equations,
 but higher derivative corrections to the equations of motion have to be introduced.
This is a  very difficult, if at all possible,
task.

The drawback of large curvature  implies that
meson masses or couplings which would be
numerical results obtained from the DBI action cannot be
considered trustworthy.
This same problem has arisen  when trying to build non-critical
holographic models. Nevertheless, Einstein-like equations
have been used to extract
qualitatively correct
results~\cite{Klebanov:2004ya,Casero:2005se,Kuperstein:2004yk}.
Considering the impressive quantitative precision
of predictions in models
like~\cite{Sakai:2004cn,SS2,Erlich:2005qh,DaRold:2005zs},
 it would be interesting to build at least phenomenological
models incorporating the tachyon physics.
Constructions like  \cite{Shock:2006qy,UmutElias} could be a
starting point.

In any case, apart from quantitative computations, we expect
the general features derived from the DBI action to hold.
Moreover, since the WZ term is topological, the results derived
from its analysis in section \ref{sect: WZ} hold even if the
curvature is large.

We end by commenting on some additional open problems.
First of all, it would be of major interest to add finite temperature
and describe the deconfinement phase transition.
The physics of fundamental matter in this regime is very
rich and
has been studied holographically (see \cite{Mateos:2007vn}
and references therein). It may be possible to generalize the
analysis to the kind of setup described in this paper.
It would also be nice to understand the physics of rotating
strings and Wilson loops.
Another possible generalization is to introduce a large number
of flavors $N_f \sim N_c$, to go beyond the quenched approximation. This may
be done along the lines of
\cite{Klebanov:2004ya,BCCKP,Casero:2006pt}.
In fact, the general form of our expressions should guarantee that they
can account for any such backreaction, as long as the theory with
fundamentals is still confining.
Finally, it would also be interesting to make contact with holographic
${\cal N}=1$ theories built on the cigar, where some exact world-sheet
computations can be done  despite the large  curvature,~\cite{cigar}.

 \addcontentsline{toc}{section}{Acknowledgments}
\acknowledgments

\noindent It is a pleasure to thank M. Berg,
A. Cotrone, U. Gursoy, D. Mateos, V. Niarchos, F. Nitti,
A. Pomarol, A.V. Ramallo and J. Sonnenschein.
RC and AP are supported by European Commission Marie Curie
Postdoctoral Fellowships, under contracts MEIF-CT-2005-024710
and MEIF-CT-2005-023373, respectively.
This work was also partially supported by
INTAS grant, 03-51-6346, RTN contracts MRTN-CT-2004-005104 and
MRTN-CT-2004-503369, CNRS PICS \#~2530 and 3059,
 and by a European Union Excellence Grant,
MEXT-CT-2003-509661.

\newpage

\appendix

\addcontentsline{toc}{section}{Appendices}
\setcounter{equation}{0}
\renewcommand{\theequation}{\Alph{section}.\arabic{equation}}
\section*{APPENDIX}

\section{Conventions} \label{app: conventions}

For the $U(N_f)$ generators $\lambda^A$, $A=0,\ldots N_f^2-1$, we take
\be\label{norm gen}
(\lambda^A)^\dagger = \lambda^A \qquad\qquad \tr ( \lambda^A \lambda^B ) =\frac{1}{2}\delta^{AB}
\ee
This in particular fixes the normalization of the $U(1)$ generator:
$\lambda^0=\frac{1}{\sqrt{2N_f}}\II$, where $\II$ is the $N_f \times N_f$ identity matrix.
For the non-abelian $SU(N_f)$ generators $\lambda^a$, $a=1,\ldots, N_f^2-1$, we have
\be
[\lambda^a,\lambda^b]=if^{ab}_{\phantom{ab}c}\, \lambda^c \qquad\qquad  \tr(\lambda^a\{\lambda^b,\lambda^c\})= d^{abc}
\ee
where $f^{ab}_{\phantom{ab}c}$ and $d^{abc}$ are, respectively,  the structure constants and  the normalized anomaly Casimir for $SU(N_f)$. Because of (\ref{norm gen}) $f^{ab}_{\phantom{ab}c}$ and $d^{abc}$ are real numbers.

We define gauge fields to be hermitian
\be\label{A decomp}
A_\mu=A^A\lambda^A=A^{U(1)}_\mu\II + A_\mu^a \lambda^a
\ee
In differential form notation, the field strength and covariant derivative read then
\be
F=dA -i A\wedge A\qquad \qquad D\equiv d - i A \cdot
\ee
where $A\cdot$ indicates the representation-dependent action of the
gauge algebra. In particular the Bianchi identity reads $DF=0$, and
the covariant derivative of the tachyon is given by
\be
DT=dT+i \, T A_L -i A_R T \qquad \qquad DT^\dagger =dT^\dagger
-iA_L T^\dagger + i \, T^\dagger A_R
\ee

Under gauge transformations,  the left and right gauge potentials, and tachyon transform in the following way
\be
\begin{split}
&A_L \to V_LA_LV^{\dagger}_L -idV_LV_L^{\dagger} \sp A_R\to V_RA_RV^{\dagger}_R -idV_RV_R^{\dagger}\\[3pt]
&T\to V_R TV_L^{\dagger}  \sp
T^\dagger\to V_L T^\dagger V_R^{\dagger}\sp V_LV_L^{\dagger}=V_RV_R^{\dagger}
=\II
\end{split}
\label{gaugevar}
\ee
An infinitesimal gauge transformation is defined as $V_\epsilon (x)=e^{\epsilon \Lambda(x)
}\simeq 1+\epsilon\, \Lambda(x)$ and the gauge transformation of a field as
$A\to A + \epsilon \delta_\Lambda A$. From (\ref{gaugevar}) we have then:
\be\label{gaugevar inf}
\begin{split}
&\delta_\Lambda A=-i\, D\Lambda=-i\,d\Lambda +[\Lambda,A]\\
&\delta_\Lambda F =[\Lambda,F]\\
&\delta_{\Lambda_L}T=-T\Lambda_L\\
&\delta_{\Lambda_R}T=\Lambda_R T
\end{split}
\ee
Notice that the generators of gauge transformations are antihermitian. When we decompose them in
their $U(1)$ and $SU(N_f)$ parts, we will write then
\be\label{Lambda decomp}
\Lambda= i\Lambda^A\lambda^A=i\alpha \II + i\Lambda^a \lambda^a
\ee
with $\alpha$ and $\Lambda^a$ real parameters. In particular we have, from (\ref{A decomp}) and (\ref{Lambda decomp})
\be\label{gauge var A}
\delta A_\mu^{U(1)} =\partial_\mu \alpha\qquad \mathrm{and} \qquad \delta A_\mu^a=(D_\mu \Lambda)^a
\ee

Currents are defined to be hermitian. We decompose a $U(N_f)$ flavor current as

\be
J_\mu=J^0_\mu \lambda^0 +J_\mu^a\lambda^a.
\ee
This decomposition corresponds to defining
the $A^{\mathrm{th}}$ component as
\be
J_{L,R\;\mu}^A=\tr_{colors} (i\bar{q} \gamma^\mu
\frac{1\pm\gamma_5}{2}\lambda^A q).
\ee
 For the $U(1)$ component $J^0$, this would be a
 strange normalization
 \be
 J^0_{L,R\;\mu}=\frac{1}{\sqrt{2N_f}}\tr (i\bar{q} \gamma^\mu\frac{1\pm\gamma_5}{2} q).
 \ee
 We therefore define a rescaled $J^{U(1)}=\sqrt{2N_f} J^0$, such that $J$ now reads
\be
J_\mu=\frac{1}{2N_f}J^{U(1)}_\mu \II +J_\mu^a\lambda^a
\ee
Notice that the normalization of $A^{U(1)}$ in (\ref{A decomp}) has been chosen in such a
way that the boundary coupling of the current to the gauge field reads
\be
2\int d^4x\, \tr(J^\mu A_\mu) =\int d^4x \left( J^{U(1)\,\mu} A_\mu^{U(1)}  +J^{a\,\mu} A_\mu^a\right)
\ee

\section{Determination of $\langle \bar qq \rangle$
from holographic renormalization}
\label{holorenorm}

When solving the equation for the modulus of the tachyon in
asymptotically AdS space, we found that it depends on two
integration constants $m_q$ and $\sigma$,
see (\ref{UVtau}). The constant associated to the
non-normalizable mode $m_q$ can be immediately identified with
the quark mass. On the other hand, $\sigma$ is related to the
quark condensate $\langle \bar qq \rangle$ in, in principle,
a non-trivial way. Schematically,
$\langle \bar qq \rangle=- \frac{\delta S}{\delta m_q}$, where
$S$ denotes the on-shell action. However, the on-shell
action is UV divergent, and must be renormalized by adding
covariant counterterms. This is done by following the so-called
holographic renormalization procedure
(for a review, see \cite{Skenderis:2002wp}).
In the following, we adapt to our case the method of
 \cite{Karch:2005ms}, where
holographic renormalization was applied to probe flavor branes.
In fact, it will be enough for our purposes to
consider a simple case in which the scalar
$\tau$ does not depend on the
Minkowski $x^\mu$-coordinates. Also for simplicity,
we consider that the metric and dilaton only depart
 from their AdS values at an order
that does not contribute to UV divergences, {\it i.e.}
$g_{xx}(z)= R_{AdS}^2/z^2(1+{\cal O}(z^5))$,
$g_{zz}(z)= R_{AdS}^2/z^2(1+{\cal O}(z^5))$ and
$e^{-\phi_{eff}}= e^{-\phi_0} (1+{\cal O}(z^5))$.

We write the action in terms of a canonically normalized, rescaled
tachyon, as defined in (\ref{UVtau}):
\be
\tau = c\ \tau_{can}\,\,,\qquad\quad c^2 \equiv \frac{e^{\phi_0}\pi}{4T_p}
\ee
The action (with a UV cutoff $\epsilon$) in terms of these quantities reads:
\be
S_{reg}=-\frac{\pi R_{AdS}^5}{2c^2}\int d^4x
\int_\epsilon^{z_{IR}} dz \frac{1}{z^5}
e^{-c^2 \tau_{can}^2} \sqrt{1+\frac23 c^2 z^2 (\partial_z \tau_{can})^2}
\label{onshell}
\ee
In the asymptotically AdS region, the equation of motion for
$\tau_{can}$ reads:
\be
-3\tau_{can}-z^5 \partial_z (z^{-3} \partial_z \tau_{can})
+\frac{c^2}{3} z^2 \frac{(\partial_z \tau_{can})\partial_z
\left(z^2 (\partial_z \tau_{can})^2\right)}
{1+\frac23 c^2 z^2 (\partial_z \tau_{can})^2}=0
\ee
and can be solved in series for small $z$ as:
\be
\tau_{can} = z\left[ \Phi_{(0)} + \frac13 c^2 z^2 \log z \Phi_{(0)}^3
+z^2 \Phi_{(2)} + {\cal O} (z^4)\right]
\label{taucanexp}
\ee
Inserting back this result in the on-shell action, one finds UV
divergencies as $\epsilon \to 0$.
They have to be subtracted by adding the following
counterterms localized at a $z=\epsilon$ slice:
\bear
S_{ct0} &=& \frac{\pi R_{AdS}}{8c^2}
\int d^4x \sqrt{-\gamma}\,\,\rc
S_{ct1} &=&-\frac{\pi R_{AdS}}{6} \int d^4x \sqrt{-\gamma}\
\tau_{can}^2\,\,\rc
S_{ct2} &=&-\frac{\pi R_{AdS}c^2}{18}
\int d^4x \sqrt{-\gamma} \,(\log \epsilon)\,
\tau_{can}^4\,\,
\eear
where $\gamma$ is the determinant of the induced metric in the
$z=\epsilon$ slice, {\it i.e.} $\sqrt{-\gamma}=R_{AdS}^4 \epsilon^{-4}$.
As pointed out in \cite{Karch:2005ms}, one can also add a finite
counterterm:
\be
S_{ct3} = \int d^4x \,\alpha\, \sqrt{-\gamma} \ \tau_{can}^4
\ee
where $\alpha$ is some constant. It was argued in
\cite{Karch:2005ms} that different values of $\alpha$ correspond
to different renormalization schemes. Defining:
\be
S_{sub} = S_{reg} + S_{ct0}+ S_{ct1}+ S_{ct2}+ S_{ct3}
\ee
the quark condensate is given by:
\be
\langle \bar qq \rangle = \lim_{\epsilon \to 0}
\left[ - R_{AdS}^{-\frac32} \epsilon \frac{\delta S_{sub}}{\delta
\tau_{can} (\epsilon)}\right]= -\frac23 \pi R_{AdS}^\frac72 \Phi_{(2)}
+R_{AdS}^\frac52 \Phi_{(0)}^3 (\frac{c^2 \pi R_{AdS}}{3}-4\alpha)
\ee
Noticing from (\ref{UVtau}), (\ref{taucanexp}) that
$\Phi_{(0)} = R_{AdS}^{-\frac32} m_q$;
$\Phi_{(2)} = R_{AdS}^{-\frac32} \sigma$ and substituting
$R_{AdS}^2 =6\alpha' = \frac{3}{\pi}$, we find:
\be
\langle \bar qq \rangle = -2 \sigma +
\frac{\pi}{3} m_q^3 (c^2 \sqrt{\frac{ \pi }{3}}-4\alpha)
\ee
In section \ref{sect: Goldstone} we have used the value of the
condensate for small $m_q$ which is unambiguously given by:
\be
\langle \bar qq \rangle \approx - 2 \sigma \,\,,
\qquad ( m_q \to 0)
\label{sigmacond}
\ee

\section{Comments on the superconnection formalism}
\label{superappendix}

In section \ref{sect: WZ} we have worked with the WZ world-volume action
for stacks of brane-antibrane pairs and used a superconnection
formalism
which makes the notation quite compact. In this appendix we review some
definitions and  properties of this construction.
The {\it supermatrices} are $2N_f \times 2N_f$ matrices
of differential forms
$M=\left(\begin{array}{cc} A & B \\ C & D\end{array}\right)$
where the blocks $A,B,C,D$ are $N_f \times N_f$ matrices.
We deal with two kinds of supermatrices: if the
blocks in the diagonal $A,D$ are composed of odd (even) differential
forms, then, the off-diagonal ones $B,C$
consist of even (odd) forms. For
instance, ${\cal A}$
(${\cal F}$)
defined in (\ref{AFdef}) are  matrices of each kind.

The multiplication of supermatrices is defined
as \cite{KrausLarsen}:
\be
M \cdot M'= \left(\begin{array}{cc} A & B \\
C & D\end{array}\right)\cdot  \left(\begin{array}{cc}
A' & B' \\ C' & D'\end{array}\right)=
\left(\begin{array}{ccc} AA'+(-)^{c'} BC' & &
AB'+(-)^{d'} BD' \\ DC'+(-)^{a'} CA' & & DD'+(-)^{b'} CB'\end{array}\right)
\ee
where $x'$ is 0 if $X$ is a matrix of even forms or 1 if $X$
is a matrix of odd forms. The associative
property of this product can be easily checked.

The supertrace  is defined as
\be
\mathrm{Str}M=\tr
\left( \begin{array}{cc} 1 & 0\\ 0 & -1 \end{array}\right)
M \qquad \Rightarrow \qquad \mathrm{Str}
\left(\begin{array}{cc} A & B \\ C & D\end{array}\right)=\tr A
- \tr D
\label{supertrace}
\ee
It is straightforward to prove the cyclic property
of the supertrace, where the
${\cal A}$-type (${\cal F}$-type) supermatrices behave as
odd (even) forms:
\be
\mathrm{Str}({\mathfrak a}\ {\mathfrak b})=-\mathrm{Str}({\mathfrak b}\ {\mathfrak a})\,;
\quad
\mathrm{Str}({\mathfrak a}\ {\mathfrak f})=\mathrm{Str}({\mathfrak f}\ {\mathfrak a})\,;
\quad
\mathrm{Str}({\mathfrak f}\ {\mathfrak g})=\mathrm{Str}({\mathfrak g}\ {\mathfrak f})\,.
\label{cyclic}
\ee
We have denoted by ${\mathfrak a}, {\mathfrak b}$
(${\mathfrak f}, {\mathfrak g}$)
generic supermatrices of the ${\cal A}$-type (${\cal F}$-type).

We also define a {\it pseudotransposition} operation:
\be
\left(\begin{array}{cc} A & B \\ C & D\end{array}\right)^{pt}
\equiv \left(\begin{array}{cc} A^t & i C^t \\ i B^t & D^t\end{array}\right)
\label{defstrans}
\ee
where $t$ denotes usual matrix transposed. This generalizes
the usual transposition in the sense that (using the same
notation of (\ref{cyclic})):
\be
({\mathfrak a}\ {\mathfrak b})^{pt} = - {\mathfrak b}^{pt}\
{\mathfrak a}^{pt}\,;
\quad
({\mathfrak a}\ {\mathfrak f})^{pt} =  {\mathfrak f}^{pt}\
{\mathfrak a}^{pt}\,;
\quad
({\mathfrak f}\ {\mathfrak g})^{pt} =  {\mathfrak g}^{pt}\
{\mathfrak f}^{pt}\,;
\quad
\label{propstrans}
\ee
Nevertheless, unlike the usual transposition,
{\it pseudotransposing} twice does not
yield the initial matrix.
Obviously, the supertrace does not change under pseudotransposition.

\section{Gauge transformation of the 1-form $\Omega_1$}
\label{app: gauge transf}

In this appendix we show that the 1-form  $\Omega_1$ defined in (\ref{Omega1}) in section
\ref{sect: axial} transforms non-trivially only under $U(1)_A$. This is an important point
 in the understanding of the $U(1)_A$ axial anomaly and it is therefore worth being explicitly shown.
For clarity of exposition, we report here the expression (\ref{Omega1}) for $\Omega_1$
\be
\Omega_1 = e^{-\tau^2} \tr \left( i A_L -iA_R
+ (\log T - \log T^\dagger)\tau d\tau \right)
\ee
It is obvious that the first two terms transform only under $U(1)_A$ transformations.
 The remaining log terms, instead, require a little more effort. First of all notice that
 we can write a general variation of $\tr(\log T-\log T^\dagger)$ as
\be\label{delta log}
\delta\tr(\log T-\log T^\dagger)=
\tr\left(T^{-1} \delta T-(T^\dagger)^{-1} \delta T^\dagger\right)=
\frac{1}{\tau^2}\tr\left(T^\dagger \delta T-T \delta T^\dagger\right)=
\frac{2}{\tau^2}\tr\left(T^\dagger \delta T\right)
\ee
where in the second equality we used the same condition on $T$ we imposed in section  \ref{sect: axial},
 $TT^\dagger=T^\dagger T=\tau^2 \mathbb{I}_{N_f}$, from where it also follows that $T\delta T^\dagger=-
 \delta T T^\dagger$, which we used in the last equality of (\ref{delta log}).

{}From (\ref{gaugevar inf}), we may show that for a vectorial ($\Lambda_L=\Lambda_R=\Lambda_V$) gauge transformation
   $\delta_{\Lambda_V}\tr(\log T-\log T^\dagger)$ vanishes.
 For an axial infinitesimal transformation ($\Lambda_L=-\Lambda_R=\Lambda_A$), from (\ref{gaugevar inf})
 and  (\ref{delta log}), we find, instead:
\be
\delta_{\Lambda_A}\tr(\log T-\log T^\dagger)=-4\tr \Lambda_A\;.
\ee
This is clearly non-zero  for abelian $U(1)_A$ axial gauge transformations only, as we wanted to show.

\section{On the WKB approximation and linear confinement}
\label{sect:WKB}

In order to use the WKB approximation to determine the spectrum,
we  start by mapping the standard eigenvalue problem to a Schr\"odinger-like equation.
We follow \cite{Schreiber:2004ie}.
We start from an equation (notice that $\lambda_n$ is
always proportional to the square of the four-dimensional mass):
\be
-\frac{1}{\Gamma(z)}\partial_z \left[\frac{\Gamma(z)}{\Sigma^2 (z)}
\partial_z \psi (z)\right] + B(z) \psi(z) = \lambda_n \psi(z)
\label{eigenvalueproblem}
\ee
and define a new radial variable $u$ and
a rescaled wave function $\a$:
\be
\frac{du}{dz}= \Sigma(z) \,\,,\qquad \alpha (u) = \Xi (u) \psi(z(u))\,\,,
\qquad
\Xi(u)\equiv \sqrt{\frac{\Gamma(z(u))}{\Sigma(z(u))}}\,\,,
\label{prescription1}
\ee
Then,
equation (\ref{eigenvalueproblem}) is rewritten
as\footnote{In this appendix, we define $V$ as the
Schr\"odinger like potential that appears in equation
(\ref{Schro-like}). It should not be confused with
the tachyon potential used in the rest of the paper}:
\be\label{Schro-like}
-\a'' (u) + V(u) \a (u) = \lambda_n \a(u)
\ee
with:
\be
V(u) = \frac{\Xi(u)''}{\Xi(u)} + B(z(u))
\label{potWKB}
\ee
One can use the WKB approximation to estimate the mass of high excitations:
\be
\frac{d\lambda_n}{dn}=2\pi \left[ \int_{u_1}^{u_2} \frac{du}{\sqrt
{\lambda_n-V(u)}}
\right]^{-1}
\label{WKB}
\ee
where $u_1$ and $u_2$ are the classical turning points.
In \cite{Schreiber:2004ie} it was observed that, quite generically, the range in the variable $u$ is finite.
Therefore, for large $\lambda_n$, the problem is similar to an infinite potential
well: the
integral behaves as $\lambda_n^{-\frac12}$
and $\lambda_n \sim m_n^2 \sim n^2$. In
\cite{Karch:2006pv} it was remarked  that the behavior of the functions
defined above can be tuned near the infrared such
that the IR turning point in equation (\ref{WKB}) goes to
infinity in such a way that the desired relation
$m_n^2 \sim n$ is obtained.
We may naively expect that
highly excited masses should depend mainly on the UV
behavior of different functions. However,
 linear confinement emerges from the IR dynamics and the size
of highly excited hadrons grows with the excitation number. Therefore,
highly excited mesons are still affected by IR physics.

We now apply this formalism  to the vector meson
equation (\ref{eqvector}). Comparing to (\ref{eigenvalueproblem}),
we define $\lambda_n \equiv m_n^2$, $B(z)\equiv 0$,
$\Gamma(z) \equiv e^{-\tilde \phi} \tilde g_{zz}^\frac12$,
$\Sigma(z) \equiv \sqrt{\frac{\tilde g_{zz}}{g_{xx}}}$.
The change of variable  to  $u$ would
involve a complicated integral, but in order to estimate the
large excitation spectrum, it is enough to study the leading
UV and IR behavior of the functions.

Near the UV, we consider an asymptotically AdS space
$\tilde g_{zz} \approx g_{xx} \approx R_{AdS}^2/z^2$,
$e^{-\tilde \phi}\approx const$, so $\Sigma\approx 1$ and
$u_{UV} = z$. The potential (\ref{potWKB})
is $V_{UV}(u) \approx \frac{3}{4 u^2}$. Thus, for
large $\lambda_n$, the classical turning point
is  at $u_1=\sqrt{3/(4\lambda_n)}$ so $u_1$ remains finite as
$\lambda_n \to \infty$
and, as expected, the UV contribution
to the integral in (\ref{WKB}) decreases as $\lambda_n^{-\frac12}$.

The intermediate region when one cannot
apply the IR nor the UV asymptotics has finite size in the
$u$-variable so its contribution to the integral also
decreases as $\lambda_n^{-\frac12}$.

Near the IR, the vev of the open string tachyon is
diverging and $g_{xx}\approx T_{QCD}$,
$\tilde g_{zz}\approx \frac{2}{\pi} (\partial_z \tau)^2$ and
$e^{-\tilde\phi} \approx e^{-\tau^2}$. By substituting
in (\ref{prescription1}), we find that the $u$-coordinate near the
IR is just
$u_{IR}\approx \sqrt{\frac{2}{\pi\,T_{QCD}}}\tau(z)$ and the
potential is $V_{IR} \approx \left(\frac{\pi}{2}T_{QCD}
\, u\right)^2$.
The classical turning point is therefore at
$u_2 = \frac{2}{\pi T_{QCD}} \lambda_n^\frac12$. Thus,
$u_2$ grows to infinity as $\lambda_n$, unlike
the cases considered in \cite{Schreiber:2004ie}. The
IR contribution to the
integral is:
\be
\int^{u_2} \frac{du}{\sqrt{\lambda_n - \left(\frac{\pi}{2}T_{QCD}
\, u\right)^2}}
= \frac{1}{T_{QCD}}
+ \dots
\ee
where the dots stand for a piece that vanishes as
$\lambda_n \to \infty$.
Therefore, the integral in (\ref{WKB})
is dominated by the IR region and:
\be
\lim_{\lambda_n \to \infty} \frac{d\lambda_n}{dn}=2\pi T_{QCD}
\ee
which, reinserting $\lambda_n = m_n^2$ yields (\ref{linconfrel}) as
we wanted to show. For the axial vectors, (see (\ref{eqaxial})),
the $z$-dependent mass for $A_\bot$
(such that $B(z)=g_{xx} v(z)^2$)
adds an extra term to the
potential near the IR $V_{IR}=(\pi^2/4 +4) T_{QCD}^2 u^2$,
leading to (\ref{wrong}).

Figure \ref{fig:schro} depicts a sketch of
the Schr\"odinger-like potential in the $u$-variable.

\FIGURE[!h]{\epsfig{file=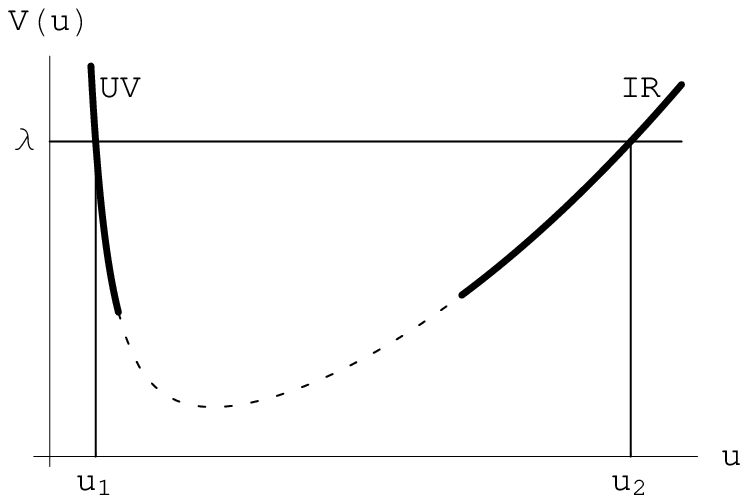,width=0.6\textwidth}
\caption{Qualitative behavior of the
Schr\"odinger-like potential
in the $u$-variable. Near the UV,
it grows as $V \propto u^{-2}$ while near the IR $V \propto u^2$.
In the middle, it may present more complicated features
which do not affect the
leading behavior of the spectrum for large excitation number.
}
\label{fig:schro}}

\end{document}